\documentclass[11pt]{article}
\usepackage[dvips]{graphicx}
\usepackage{amsfonts}
\usepackage{amssymb}
\usepackage{amsmath}
\usepackage{epsfig}
\usepackage{multirow}
\usepackage{latexsym}
\usepackage{graphicx}
\usepackage{float}
\usepackage{hyperref}

\usepackage[nosort]{cite}

\newcommand{\nn}{\nonumber}
\newcommand{\be}{\begin{equation}}
\newcommand{\ee}{\end{equation}}
\newcommand{\bea}{\begin{eqnarray}}
\newcommand{\eea}{\end{eqnarray}}

\newcommand{\half}{\frac{1}{2}}

\newcommand{\vphi}{\varphi}

\newcommand{\beqa}{\begin{eqnarray}}
\newcommand{\eeqa}{\end{eqnarray}}

\newcommand{\hx}{\hat{\xi}}

\newcommand{\f}{{\bf 5}}
\newcommand{\fb}{{\bf \bar{5}}}
\newcommand{\te}{{\bf 10}}
\newcommand{\teb}{{\bf \bar{10}}}
\newcommand{\op}{\oplus}

\topmargin
-0.7cm
\textwidth
15.5cm
\textheight
22.8cm
\oddsidemargin
0.7cm
\evensidemargin
1.2cm

\begin{document}

\makeatletter
\@addtoreset{equation}{section}
\makeatother
\renewcommand{\theequation}{\thesection.\arabic{equation}}
\pagestyle{empty}
\rightline{CPHT-RR012.0312}
\vspace{2.2cm}
\begin{center}
\Large{\bf Wavefunctions and the Point of $E_8$ in F-theory\\[12mm]}
\large{Eran Palti \\[5mm]}
\small{Centre de Physique Th´eorique, Ecole Polytechnique, CNRS, F-91128 Palaiseau, France.}\\[1mm]
\small{palti@cpht.polytechnique.fr} \\ [14mm]
\small{\bf Abstract} \\[2mm]
\end{center}
\begin{center}
\begin{minipage}[h]{16.0cm}
In F-theory GUTs interactions between fields are typically localised at points of enhanced symmetry in the internal dimensions implying that the coefficient of the associated operator can be studied using a local wavefunctions overlap calculation. Some F-theory $SU(5)$ GUT theories may exhibit a maximum symmetry enhancement at a point to $E_8$, and in this case all the operators of the theory can be associated to the same point. 
We take initial steps towards the study of operators in such theories. We calculate wavefunctions and their overlaps around a general point of enhancement and establish constraints on the local form of the fluxes.
We then apply the general results to a simple model at a point of $E_8$ enhancement and calculate some example operators such as Yukawa couplings and dimension-five couplings that can lead to proton decay.
\end{minipage}
\end{center}
\newpage
\setcounter{page}{1}
\pagestyle{plain}
\renewcommand{\thefootnote}{\arabic{footnote}}
\setcounter{footnote}{0}


\tableofcontents


\section{Introduction}
\label{sec:intro}

Given that string theory, if correct, should account for all current and future particle physics experimental data, string phenomenology is a data-rich subject. However much of this data lies not with the observed states of the Standard Model and its extensions, but with the operators of the theory. This includes measured values of operators, or couplings, in the SM as well as constraints on operators of possible extensions of it to, for example, supersymmetric or Grand Unified Theories. Quantitative calculations of operator coefficients within realistic string theory models are typically difficult because they require tools beyond algebraic geometry which are difficult to apply to complicated higher dimensional geometries with various fluxes and branes present. This means that, apart from for particularly simple geometries, it is difficult to say much beyond the bottom-up rule that if an operator is not forbidden by some symmetry then it is present with an order one coefficient. This approach can be refined by considering for example the scaling of operators with various moduli and other coarse properties of the background, and for many applications this is sufficient, nonetheless a quantitative computation is often missing. 

One class of constructions where we might be able to present a quantitative study of operators within a general setting is in type IIB or F-theory where the set of cubic normalisable operators can be calculated using wavefunction overlaps on smooth backgrounds \cite{Cremades:2004wa,Abe:2008fi,Conlon:2008qi,Heckman:2008qa,Hayashi:2009ge,Choi:2009pv,Font:2009gq,Cecotti:2009zf,Conlon:2009qq,Hayashi:2009bt,Aparicio:2011jx,Oikonomou:2011kd,Camara:2011nj}.\footnote{The singular limit of geometric IIB/F-theory constructions which are branes on singularities also allow for calculating couplings but using different techniques \cite{Aldazabal:2000sa,Conlon:2008wa,Krippendorf:2010hj,Burgess:2011zv,Dolan:2011qu}.} This by itself is still a task beyond our reach for a general background, but there is an additional advantage which is that the couplings are naturally associated to only a relatively small patch within the full Calabi-Yau. This follows for the common constructions where gauge fields are associated to divisors wrapped by 7-branes, matter fields to curves where 2 branes intersect, and cubic couplings to points where 3 curves intersect \cite{Vafa:1996xn,Morrison:1996na,Donagi:2008ca,Beasley:2008dc,Beasley:2008kw,Donagi:2008kj,Denef:2008wq}. Given this we can study the couplings by calculating the wavefunctions and their overlaps on a local flat patch around a point of interaction. This was done for Yukawa couplings in \cite{Heckman:2008qa,Font:2009gq,Conlon:2009qq,Aparicio:2011jx} and for some other couplings, particularly to heavy modes, in \cite{Camara:2011nj}. 

There is no doubt that one of the most attractive aspects of IIB and F-theory models is the control over the full global setup. They are the constructions where the vacuum structure and its stability can be argued to be best understood \cite{Giddings:2001yu}, and there has been substantial development in understanding of global model building in F-theory.\footnote{See \cite{Denef:2008wq} for reviews and \cite{Grimm:2010ez,Grimm:2011tb,Braun:2011zm,Marsano:2011hv,Krause:2011xj,Grimm:2011fx,Krause:2012yh} for some of the most recent developments.} However it is also fair to say that one of the deepest problems facing string phenomenology is that of the landscape and in particular how to extract semi-universal phenomenological aspects. A major advantage of the local approach in this respect is that the large parameter space of the CY geometry and the many fluxes gets pulled back, or projected, to just a small set of adjustable local parameters around the point of interaction. This is particularly appealing in the sense that although the large number of vacuum configurations all necessarily affect the vacuum energy, or cosmological constant, the projection to the local patch implies that the various configurations can give rise to a comparably small set of particle physics models. We can then hope to use this and the additional calculability of local models to impose a large number of constraints, corresponding to the many operator coefficients, on a relatively small number of local parameters. 

An important disadvantage of studying the theory around a point of interaction is that the local fluxes and geometry at different points, say at an up-type Yukawa point and a down-type Yukawa point, are not related in a simple way and so it is difficult to calculate relations between the two, such as the CKM matrix. It is therefore interesting to consider models where both the Yukawa points are located in the same local patch. Indeed if the observed flavour structure of the SM, which has strong correlations between the up and down sector, has a geometric origin it is natural to expect that the two points should be experiencing the same geometry. Since the interaction points are associated to an enhanced local symmetry, for example $SO(12)$ and $E_6$ for $SU(5)$-GUT Yukawas, their coincidence corresponds to a further enhancement in the symmetry. For the Yukawas this amounts to at least $E_7$ but if we consider more interactions it is natural to consider the maximum enhancement to $E_8$. Indeed this reasoning was the motivation presented for studying the idea of a point of $E_8$ enhanced symmetry in \cite{Heckman:2009mn}. Such a construction enjoys this motivation, but a further slightly different motivation is that in the spirit of being able to calculate operators on a local patch we are motivated to study the richest and most complete theories to which these calculations apply. This way we are able to calculate many operators that in models where the interactions are spread out throughout the CY we would not be able to. In some ways these are the simplest realistic theories where we can hope to calculate the bulk of the operators as a function of a relatively small set of parameters.

The work in this paper is motivated by the idea of a highly calculable, quite general, theory around a point of $E_8$. The idea being to take the point of $E_8$, or some other enhanced symmetry, as a starting point and calculate the couplings of the fields to each other as a function of the local values of fluxes and geometry. There are GUT singlets typically present at such points which parameterise deformations of it, and therefore these theories also include the set of theories where the singlets have a relatively small vev, or where the point of $E_8$ is slightly deformed. In principle such theories are rich enough to have substantial applications to phenomenology.\footnote{However there are two effects which may, but not must, play an important role in phenomenology which we do not consider, these are local monodromies \cite{Hayashi:2009ge,Cecotti:2010bp} and possibly non-commutative deformation associated with flux or non-perturbative effects \cite{Cecotti:2009zf,Conlon:2009qq,Marchesano:2009rz,Aparicio:2011jx}. Calculating the operators in a setup including these effects most likely will require numerical analysis which we leave for future work and in this paper we will consider as a starting point the most basic scenario where these effects are absent.}

Aside from applications to models based on the point of $E_8$ we will study quite general properties of the effective gauge theory and its wavefunction solutions. Indeed the bulk of the paper will not specify a particular gauge group but solve for the most general local forms of the wavefunctions for arbitrary gauge flux and Higgs backgrounds, in the leading linear approximation. Once these are obtained we will apply them to a toy model around a point of $E_8$. The paper is set out as follows. In section \ref{sec:effthe} we will set up the general formalism that we will use. We begin by introducing the effective theory and discuss in detail constraints required for its validity. We then study quite generally the relation between the local flux and Higgs values, and the form of the wavefunctions and what type of chiral matter they correspond to, thereby developing the local relation between flux and chirality which plays an important role subsequently. In section \ref{sec:wave} we present explicit solutions for the form of the wavefunctions, for both massless and massive modes, and their overlaps for quite general flux and Higgs backgrounds. In section \ref{sec:modbui} we utilise the results developed in the previous sections to study a toy model based on a point of $E_8$ enhancement. Within this model we calculate some phenomenologically interesting quantities like the Yukawa couplings, and various higher dimension operators particularly those associated to dimension-five proton decay. We summarise the work in section \ref{sec:summary}. In the appendix we present a semi-local analysis of the flux and chiral spectrum in the absence of any monodromies, so for a fully split spectral cover.

\section{The effective theory}
\label{sec:effthe}

The setup we are studying is of a 7-brane in F-theory, wrapping a four-cycle $S$, carrying a gauge group $G_S$ and exhibiting an enhancement at a point on its worldvolume to a gauge group $G_P$ which is at least 2 ranks higher. In the upcoming sections we will keep all the formalism general so that it can be applied to any 7-brane gauge group enhancing up to any higher group. However for concreteness we will refer to the particular case of $G_S=SU(5)$ enhancing to $G_P=E_8$, and in section \ref{sec:modbui} we will consider models that are specific to this choice. In the infrared the theory on the 7-brane is described by a twisted 8-dimensional $\mathcal{N}=1$ gauge theory, with gauge group $G$ and support on $\mathbb{R}^{1,3}\times S$, where $S$ is a 4-dimensional K\"ahler sub-manifold of the F-theory Calabi-Yau fourfold $X$. The twisting accounts for the embedding of the 7-brane into the four-fold in a supersymmetric way \cite{Donagi:2008ca, Beasley:2008dc}. The field content of the theory can be written in terms of 4-dimensional $\mathcal{N}=1$ multiplets as
\bea
{\bf A}_{\bar{m}} &=& \left\{ A_{\bar{m}}, \psi_{\bar{m}}, \mathcal{G}_{\bar{m}}\right\} \;, \label{hol}\\
{\bf \Phi}_{mn} &=& \left\{ \left(\varphi_H\right)_{mn}, \chi_{mn}, \mathcal{H}_{mn} \right\} \;, \nn\\
{\bf V} &=& \left\{ \eta, A_{\mu}, \mathcal{D} \right\} \;,\nn
\eea
where by abuse of notation we denote here and henceforth both the 8-dimensional fields and their internal wavefunction profiles by the same symbols and will only make an explicit distinction when necessary.
The subindices on the fields denote their local differential structure on $S$ and so take values $m=1,2$. Here ${\bf A}$ and ${\bf \Phi}$ are chiral multiplets with respective F-terms $\mathcal{G}$ and $\mathcal{H}$. ${\bf V}$ is a vector multiplet with D-term $\mathcal{D}$. $A_{\bar{m}}$ and $\left(\varphi_H\right)_{mn}$ are complex scalars while $\psi_{\bar{m}}$, $\chi_{mn}$, and $\eta$ are fermions.

Following the ideas presented in \cite{Morrison:1996na,Beasley:2008dc} we can model the intersecting brane combinations that lead to an enhancement to $E_8$ at a point in $S$ by considering the 8-dimensional gauge theory to have the gauge group $G=E_8$. Within this theory we give the scalar Higgs component $\left(\varphi_H\right)_{mn}$ a spatially varying vacuum expectation value (vev) which, at a generic point on $S$ breaks the gauge group to $SU(5)_{GUT}\times U(1)^4$. The Higgs vev vanishes on certain curves and points in $S$ on which therefore the gauge group enhances. Relevant to us will be curves where the Higgs vev vanishes along one $U(1)$ factor inducing an enhancement of $SU(5)$ by rank 1 to $SU(6)$ or $SO(10)$. As well as a point where the full Higgs vev vanishes inducing an enhancement by rank 4 to $E_8$. 

The equations of motion of the theory on $\mathbb{R}^{1,3}\times S$ were derived in \cite{Donagi:2008ca, Beasley:2008dc, Conlon:2009qq}. Solving them on some complicated background geometry $S$ is a prohibitively difficult task. However since our primary interest is in the region around a single point, of $E_8$ enhancement, in $S$ we can approximate the theory locally as flat space $S \sim {\mathbb C}_2$. Within this approximation the equations simplify considerably. 

Having described the general setup we go on to specify the details of the model. These are similar to, but more general than, the setup discussed in \cite{Camara:2011nj} and the related configurations of \cite{Marchesano:2008rg,Camara:2009xy,Marchesano:2010bs,Aparicio:2011jx}, to which we refer for further details on some of the concepts introduced below. 

\subsection{The equations of motion}
\label{sec:eqmot}

We are interested in the equations of motion for fluctuations of the fields about a background with a Higgs vev and non-vanishing gauge field flux along $S$. Locally, the leading order contributions from the Higgs and gauge fields are linear in the local co-ordinates $z_1$ and $z_2$ and so take the form
\be
\langle\varphi_H\rangle=M_{K} R\, m^a_{i} z_i \, Q_a dz_1\wedge dz_2+\ldots\ ,\label{phi}
\ee
\be
\langle A\rangle =-M_{K}\,\textrm{Im}( M^a_{ij} z_id\bar z_j)Q_a + \ldots\ , \label{a} 
\ee
where the dots denote higher order terms in the two local complex coordinates $z_1$ and $z_2$. Here the $m^a_i$ and $M^a_{ij}$ are complex constants which, in the case of $E_8$ enhancement, denote the vevs along the 4 $U(1)$ factors with associated generators $Q_a$. Actually it is more convenient to turn on the vev within $S(U(1)^5)$ so that $a=1,..,5$ but we have to impose an additional tracelessness constraint
\be
\sum_{a=1}^5 m^a_i = \sum_{a=1}^5 M^a_{ij} = 0 \;.
\ee
We will also allow for flux $M^Y_{ij}$ along the Hypercharge $U(1)$ inside $SU(5)_{GUT}$ with generator $Q_Y$. 
Note that the local expansion of the Higgs and the flux begins with a linear term in the $z_i$ and there is no constant term. For the Higgs background this amounts to defining the enhancement point to be at the origin. For the gauge field a constant term locally is pure gauge and so can be gauged away.

The mass scale $M_{K}$ is a {\it local} mass scale which involves the cutoff scale of the theory $M_*$ and scales with a homogeneous rescaling of the local metric by a length scale $R_{\parallel}$ as 
\be
M_K = \frac{M_*}{R_{\parallel}} \;.
\ee 
The dimensionless scale $R$ is associated to the {\it local} normal length scale $R_{\perp}$ to $S$ and scales as 
\be
R \equiv R_{\parallel}R_{\perp} \;.
\ee
The scaling of the Higgs with $R_{\perp}$ follows from the pullback of the normal metric component to the world-volume of the 7-brane as discussed in \cite{Camara:2011nj}. The gauge field is real and leads to flux of complex type $(1,1)$ and the Higgs is holomorphic, these two requirements amounting to the preservation of half the supersymmetry by each one. Further, given the local flat metric with associated Kahler form 
\be
\omega = \frac{i}{2} \left(dz_1 \wedge d\bar{z}_{\bar{1}} + dz_2 \wedge d\bar{z}_{\bar{2}} \right) \;, \label{localkahler}
\ee
the D-term constraint, $\omega \wedge F = 0$, imposes that
\be
\mathrm{Re}\;M^a_{11} = -  \mathrm{Re}\;M_{22}^a \;.
\ee

Note that the flux associated to the gauge field (\ref{a}) reads
\be
F = \frac{i}{2} M_{K}^2 \left(M_{ij}+\bar{M}_{ji}\right) dz_i \wedge d\bar{z}_j \label{aflux} \;.
\ee
Not all the flux parameters appearing in (\ref{a}) also appear in (\ref{aflux}): for example only the real part of $M_{11}$ and $M_{22}$ appear in the flux. However the wavefunction calculation does depend on the choice of gauge field and so these parameters do feature in some of the calculation.

With this background the equations of motion for fermionic fluctuations which lead to massless four-dimensional fields read \cite{Donagi:2008ca, Beasley:2008dc, Conlon:2009qq,Marchesano:2010bs,Aparicio:2011jx,Camara:2011nj}\footnote{Note that we have slightly changed conventions from \cite{Camara:2011nj} to match those of \cite{Aparicio:2011jx} by flipping the sign of some components in $\mathbb{D^{\pm}}$.}
\begin{equation}
\mathbb{D^-}\Psi=0 \label{fterm} \;,
\end{equation}
with,
\begin{equation}
\mathbb{D^{\pm}}=\begin{pmatrix}0& D_1^{\pm}& D_2^{\pm} & D_3^{\pm}\\
-D_1^{\pm}& 0& -D_3^{\mp}& D_2^{\mp}\\
-D_2^{\pm}& D_3^{\mp}&0&-D_1^{\mp}\\
-D_3^{\pm}& -D_2^{\mp}&D_1^{\mp}&0 \end{pmatrix}\ , \qquad \Psi=\begin{pmatrix}\eta\\ \psi_{\bar 1}\\ \psi_{\bar 2}\\ \chi\end{pmatrix} \;,
\label{z2}
\end{equation}
and
\begin{align}
D_i^-&\equiv \partial_i-\frac{1}{2}q_a\bar{M}^a_{ji}\bar{z}_j& D_i^{+}&\equiv \bar \partial_i+\frac{1}{2}q_aM^a_{ji}z_j\, \qquad i=1,2\label{gaugecov}\\
D_3^-&\equiv - R \, q_a \bar{m}^a_{i} \bar{z}_i & D_3^+&\equiv R \, q_a m^a_{i} z_i \;.
\end{align}
Here $q_a$ denote the charges of the fields under the generator $Q_a$. Since the charges are always contracted with the Higgs or fluxes it is convenient to introduce the notation
\be
M_{ij} \equiv q_a M_{ij}^a \;,\;\; m_i \equiv q_a m_i^a \;. \label{efffluxhiggs}
\ee

It is useful to consider the 8-dimensional fields as a vector $\Psi$ which contains the fermionic components of a single ${\cal N}=4$ multiplet. The reason is that four-dimensional mass eigenstates will in general be combinations of the 8-dimensional fields. The particular splitting (\ref{hol}) is rather more convenient for writing down the equations in coordinates. The equations of motion for fluctuations which lead to massive four-dimensional fields can be written as \cite{Marchesano:2010bs,Aparicio:2011jx,Camara:2011nj}
\be
\mathbb{D}^+\mathbb{D}^-\Psi=\left|m_{\lambda}\right|^2 \Psi \label{masseom} \;,
\ee
with associated four-dimensional mass $M_K m_{\lambda}$.

Having described the effective theory we discuss the range of parameters for which the theory is valid. This was studied in \cite{Camara:2011nj} which we summarise and expand upon. There are a number of approximations being used and it is worth splitting the associated constraints. Firstly we have the constraint that the metric takes the flat form (\ref{localkahler}). Note that with our conventions we first write it as a flat space metric and then rescale away the factors of $R_{\parallel}$ by writing in terms of a vielbein basis. This means that corrections to the metric are suppressed within the region defining our local patch
\be
\mathrm{Patch:}\; \left|z_i\right| \ll 1 \;.
\ee
Outside this patch we can not trust the form of the equations and their associated wavefunction solutions. 

Even within the local patch we may have corrections to the theory itself. These come from large localised energy densities which are associated to derivatives of the Higgs and background gauge field and so to keep these small we require
\bea
\frac{M_{ij}}{R_{\parallel}^2} \ll 1 \;, \;\;
\frac{m_{i}R_{\perp}}{R_{\parallel}} \ll 1 \;. \label{fluhigcor}
\eea
These amount to dilute fluxes and small brane intersection angles and if they are violated we expect corrections to the theory coming from higher derivative operators.


The full constraints are most simply satisfied by taking 
\be
R_{\parallel} \gg R_{\perp} \gg 1 \;. \label{localscalesize}
\ee
As discussed, it is important to keep in mind that all of the parameters, $R_{\parallel}$, $R_{\perp}$, $m_i$ and $M_{ij}$ are {\it local} parameters and so are not directly tied to their associated global scales. Even so, at least in the local patch the manifold should be near homogeneous, and if this near homogeneity is extended to the whole of $S_{GUT}$ and the CY then we have that
\bea
R_{\parallel} &\sim& \mathrm{Vol}\left(S_{GUT}\right)^{\frac14} \sim \alpha_{GUT}^{-\frac14} \;, \\
R_{\perp} &\sim& \left(\frac{M_{\mathrm{Planck}}}{M_*}\right)^p \;.
\eea
where $p$ is some power which varies according to the embedding of $S_{GUT}$ into the CY. For example $p=1$ for a torus and $1/3$ for a contractible $S_{GUT}$. 

Given this there are a number of important observations. The first is that the decoupling limit: $M_{Planck} \rightarrow \infty$ with $\alpha_{GUT} \rightarrow \mathrm{finite}$, is not naturally compatible with the constraint coming from small Higgs variations (\ref{fluhigcor}). This is because the parameters $m_i$, like the fluxes $M_{ij}$, are quantised in a global context. Physically this issue amounts to the fact that if the $U(1)$ branes go into the bulk of the CY they can not have small intersection angles with the GUT brane throughout the surface $S_{GUT}$. Intuitively this can be seen by noting that the Higgs field measures the distance to the bulk branes and so in going around a loop in $S_{GUT}$, of typical length $R_{\parallel}$, it must range all the way to the global scale $R_{\perp}$ and back again and therefore must have a large gradient and so large intersection angles. In this case the approach of modeling the intersecting brane setup using an 8-dimensional theory with a higher gauge group which is broken by the Higgs field breaks down. Rather the appropriate model would have to be 10-dimensional with a product gauge group treating the bulk branes as separate.\footnote{A more field-theory perspective on this issue can be seen by thinking about the energy scale at which the theory enhances to the higher gauge group all over $S_{GUT}$. This energy scale goes like the bulk scale, for a non-compact setup diverges, and in particular is higher than the local cutoff scale of the theory and so such a theory based on an enhanced group is inconsistent.}

However it is important to stress that this does not imply a breakdown of the theory for the actual finite value of $M_{Planck}$. Rather it requires some degree of inhomogeneity in the CY so that the local $R_{\perp}$ in patch differs enough from the global value which gravity feels. In fact this is another motivation for considering the point of $E_8$ since then all the relevant calculations depend on the local value of $R_{\perp}$. Another possibility is that the intersecting brane configuration is such that the $U(1)$ bulk branes also wrap local contractible cycles so that the Higgs does not range over the full CY volume.

Another consequence of extending the local scale $R_{\parallel}$ to $S_{GUT}$ is that the local approximation, where the wavefunctions are localised on a small patch within $S_{GUT}$ restricts the parameter space significantly. This is because of the relatively strong expected coupling at the GUT scale so that $R_{\parallel} \sim \alpha_{GUT}^{-\frac14} \sim {\cal O}\left(1\right)$ is just not that large (note that additional massive fields which typically increase the GUT coupling make this situation even worse). To localise wavefunctions within patches much smaller than this requires strong energy densities for which our effective theory would break down. This means that the wavefunctions sometimes still have a significant tail at distances $\left|z_i\right| \sim 1$ from the point of $E_8$ and so for those wavefunctions the accuracy of the overlap calculation is reduced. Practically however these corrections can be kept well under control. On the other hand, as discussed in the introduction,  the smallness of $S_{GUT}$ implies that even for general F-theory models which do not have a point of $E_8$ enhancement, since the wavefunctions on $S_{GUT}$ may have significant overlap, a model based on a point of $E_8$ may be a decent approximation.
In the rest of the analysis we can consider working in the $\alpha_{GUT} \rightarrow 0$ limit where the results discussed and the localisation (onto matter curves) approximations become exact. 

Just as the Higgs field induces localisation onto the matter curves the flux on the matter curves induces localisation within the curves. However this latter effect does not become exact in the limit $\alpha_{GUT} \rightarrow 0$ but rather the degree of localisation is controlled directly by the flux $M$. Since this is expected to be of order one this localisation is typically less strong than that transverse to the matter curve and forms again a source of possible corrections to the theory coming from the wavefunctions sensing the curvature corrections of the metric on the matter curve. The localisation along the curves is not so important for calculating the operators of the theory since the appropriate integrals are fully localised within $S_{GUT}$ by the Higgs vev. However it does feature in the normalisation of the wavefunctions and in determining the local chiral spectrum of the theory. We discuss this in more detail in the next section.

\subsection{Local and global modes}
\label{sec:locspe}

The key objects in calculating the operators of the four-dimensional theory are the internal wavefunctions of the four-dimensional fields along $S_{GUT}$. The background Higgs profile (\ref{phi}) ensures that the wavefunctions are localised onto matter curves. The profiles of the wavefunctions along the curves themselves are more complicated and are primarily sensitive to the background flux (\ref{a}). As well as controlling the wavefunction profiles along the matter curves the flux also generates chirality in the massless spectrum. The interplay between these two effects of the flux plays an important role in the spectrum analysis and particularly so for the case of a point of $E_8$. In this section we study and develop this connection as well as elucidating more generally the question of how much can we learn about the wavefunctions from a given local patch along the matter curve. 

\subsubsection{A global example}
\label{sec:globex}

As a reference to the general concepts we will discuss it is useful to have an explicit example. In this subsection we briefly study some properties of wavefunctions over an explicit matter curve ${\cal C}$ which we take to be complex projective space ${\mathbb P}^1$. This model is slightly different from our model in that it is 1-complex dimensional and untwisted, where as in our model wavefunctions are localised onto 1-complex dimensional curves inside a 2-complex dimensional twisted theory. Nonetheless the essential features will be similar.

We begin by briefly recalling the most common way to count chirality on ${\mathbb P}^1$ with background flux. We specify the background flux through the transition functions of the line bundle which for homogeneous coordinates $Z_i$, and integer flux $M$, go as $\left(Z_i^a/Z_i^b\right)^{\left(M-1\right)}$, with patches labeled by $a$ and $b$. Therefore global holomorphic sections are counted by homogeneous polynomials of degree $M-1$ of which there are $M$. Then, since we are counting solutions to the F-term equations only, we can go to the holomorphic gauge for which the Dirac operator reduces to the simple Dolbeault operator and so solutions are given by any holomorphic sections. There are $M$ of these and so there are $M$ massless modes, which are chiral since the conjugate partners would correspond to anti-holomorphic sections of which there are none because of the holomorphic line bundle transition functions.

This way of counting chirality is intrinsic to the holomorphic gauge. However the holomorphic gauge is not physical and is only useful for understanding rather coarse properties. For example it does not lead to the physical couplings which have non-holomorphic components and it is not useful for understanding the massive spectrum. In this paper we work in the unitary gauge and the aim of this subsection is to show how chirality can be understood in this gauge.

Let us denote the affine complex coordinate on ${\cal C}$ as $z$ and put the canonical Fubini-Study metric on it
\be
ds^2 = \frac{dzd\bar{z}}{\left(1+z\bar{z}\right)^2} \;.
\ee
We now wish to turn on constant flux through ${\cal C}$ and calculate the wavefunction solutions of the Dirac equation. This was studied in detail in \cite{Conlon:2008qi} whose results we use. The flux takes the form
\be
F = \frac{-iM}{\left(1+z\bar{z}\right)^2} dz \wedge d\bar{z} \;,
\ee
with associated gauge field
\be
A = \frac{iM\bar{z}}{2\left(1+z\bar{z}\right)} dz + \mathrm{c.c.} 
\ee
Turning on flux along the matter curve is known to induce chirality in the massless spectrum and the net chiral index is given by the integral of the flux over the curve which in this case is
\be
\chi_{\mathrm{global}} = -\frac{1}{2\pi}\int_{\cal C} F = M \;. \label{globch}
\ee
We are particularly interested in how this chirality manifests itself in the wavefunctions. The massless wavefunction solutions are \cite{Conlon:2008qi}
\bea
\psi_- &=& f_-(\bar{z}) \left(1+z\bar{z}\right)^{\frac{1-M}{2}}\;, \nn \\
\psi_+ &=& f_+(z)\left(1+z\bar{z}\right)^{\frac{1+M}{2}}\;. \label{wavep1}
\eea
The two wavefunction solutions are for fields of opposite charges and are the two chiralities. The holomorphic and anti-holomorphic functions $f_+(z)$ and $f_-(\bar{z})$ are general functions, which we can take to be polynomials, and therefore there are an (equal and) infinite number of solutions to the Dirac equation for each chirality. However chirality arises upon requiring that physical solutions are normalisable so that 
\be
\int_{\cal C} dzd\bar{z}\sqrt{g} \psi_{\pm}^{\dagger} \psi_{\pm} = \int_{\cal C} dzd\bar{z}\frac{|f_{\pm}|^2}{\left(1+z\bar{z}\right)^{(1\mp M)}}\;,
\ee
is finite. Now say $M>0$, then there are no normalisable solutions for $\psi_+$ while for $\psi_-$ there are $M$ independent solutions corresponding to $f_-(\bar{z})$ being anti-holomorphic polynomials of degree up to $M-1$. The opposite is true for $M<0$ while for $M=0$ there are no normalisable modes. This is exactly the results of the chirality formula (\ref{globch}).

Now the question we are interested in is how much of the geometry of ${\cal C}$ did we need to know to get these results? The chirality is determined by the convergence properties of $zg^{\frac14}\psi_{\pm}$ as $z \rightarrow \infty$. Say $f_{\pm}$ are taken as polynomials of order $p$, then the crucial object is the existence and position of the turning point in the function
\be
\left|zg^{\frac14}\psi_{\pm} \right| = |z|^{p+1}\left(1+|z|^2\right)^{(-1\pm M)/2}  \;,
\ee
which occurs at
\be
\left|z_{\mathrm{turning}}\right| = \sqrt{\frac{p+1}{\mp M - p}} \;.
\ee
The existence of this turning point holds the same information as the chirality formula (\ref{globch}) by counting the possible positive integer values of $p$. For $M \gg p$ we have $\left|z_{\mathrm{turning}}\right| \rightarrow 0$ while for $M=p$ we have $\left|z_{\mathrm{turning}}\right| \rightarrow \infty$. Therefore for large $p$ we need to know the full geometry of ${\cal C}$ in order to count the full chirality induced by $M$. However for the chiral states with $p<M$ we do not need to know the full geometry and for large $M/p$ the existence of a turning point can be studied for values of $z \ll 1$ and therefore a flat space expansion is sufficient to identify it. Each of the turning points can be associated to a turning point in the wavefunctions.\footnote{With the exception of the limiting case $p=M-1$ where the wavefunctions are a constant, though this exception is not present in the presence of twisting.} 
The extreme case being $p=0$ which is associated to a maximum of the wavefunction at the origin. Let us see this explicitly: in the local patch the wavefunctions (\ref{wavep1}) change form and rather take the form of Gaussian exponentials\footnote{The extra factor of $\frac12$ with respect to the form of our local wavefunctions in the exponent is because we have not included the twisting of the theory.}
\bea
\left.\psi_-\right|_{\mathrm{local}} &=& f_-(\bar{z}) e^{\frac{-M+1}{2}|z|^2} + \ldots \;, \nn \\
\left.\psi_+\right|_{\mathrm{local}} &=& f_+(z) e^{\frac{M+1}{2}|z|^2} + \ldots\;. \label{wavep1loc}
\eea
Now with the form (\ref{wavep1loc}) we directly see which wavefunctions are normalisable or not. However it seems that there are an infinite number of normalisable solutions since any polynomial of degree $p$ would still be dominated by the exponential for large $|z|$. This is simply an artifact of the fact that for large enough $p$ the turning point where the exponential dominates the polynomial is forced further away from the origin and our perturbative expansion breaks down. Nonetheless the presence of a turning point and the associated local chirality can be established within the local flat patch for the wavefunctions with small $p$.

\subsubsection{Local Wilson lines and chirality}
\label{sec:locchir}

In the previous section we considered an explicit global example of a matter curve. We showed that both chiralities solve the Dirac equation and that chirality can be understood to arise from the normalisability of the wavefunctions. We then showed that in the example this was directly related to the presence of a single peak in the wavefunctions, and that chirality was counted by the number of independent wavefunctions with different peak positions. 

Consider now only working in a local patch around a point in the matter curve. This is our setup because we must define our local patch around a particular point of interaction which is where the matter curve intersects other curves. The position of this point on the curve must be kept arbitrary in a local approach since it is only determined globally. As we have seen above chirality is determined by the normalisability of the wavefunctions which generally requires global knowledge. However we have also argued that the chirality is related to local turning points. We showed this explicitly for the example in section \ref{sec:globex} and also showed that within this approach it is true that determining the full spectrum requires global knowledge but identifying a single massless mode can be done locally.\footnote{There may be a way to work from a local picture to a global one where identifying a wavefunction turning point can be associated to a global chiral mode perhaps using the approach to Morse theory introduced in \cite{Witten:1982im}. Though it is not clear to the author how, or if it is possible, to formally apply these techniques to matter curves with flux and show generally an analogous relation between the local turning points in wavefunction solution and global chiral zero modes.}

With these ideas in mind we can classify two types of zero modes. We fix the global boundary conditions and wavefunction basis and specify some point $q$ on the curve where the interaction is localised. Then we define {\it global modes} as ones where the wavefunction peak is not located within the local patch around $q$, and {\it local modes} as ones whose wavefunction peaks within the patch. With these definitions, generally we can not ascertain information regarding the normalisability of the wavefunctions for global modes in the local patch and in particular make any statements regarding the chirality of the associated mode. Conversely, the definition of local modes is a chiral one: only one chirality can have a wavefunction peak in a given local patch.

Let us now return to our general local $E_8$ gauge theory. We seek a formula that will determine the chirality of local modes. This should depend on the local flux values as well as the local Higgs values. For the background Higgs and flux values (\ref{phi}) and (\ref{a}), this formula takes the form
\be
\chi_{\mathrm{local}}\left(q^a\right) = -\mathrm{Re}\left[\left(M_{12}+\bar{M}_{21}\right)\bar{m}_1 m_2 + M_{11} \left(\left|m_1\right|^2-\left|m_2\right|^2\right)\right] \;. \label{locchi}
\ee
For a given state with charges $q^a$, if $\chi_{\mathrm{local}}>0$ there is a localised chiral mode. In application to model building the formula (\ref{locchi}) presents non-trivial constraints on the possible local massless spectrum. The formula can be most simply derived by explicitly evaluating the local intersection number. The Poinca\'{e} dual two-form to the matter curve can be constructed through the vanishing locus of the Higgs background as given in (\ref{phi})
\be
\mathrm{PD}\left[ {\cal C} \right] = \partial\left<\vphi_H\right> \wedge \bar{\partial}\left<\bar{\vphi}_H\right>\;.
\ee
Then intersecting this with the flux (\ref{aflux}) gives 
\be
F \wedge \mathrm{PD}\left[ {\cal C} \right] = \chi_{\mathrm{local}}\; \eta_4\;,
\ee
where $\eta_4$ is the volume form on the local patch. The massless and massive spectrum of local modes, as well as their wavefunctions, completely depends on the sign of $\chi_{\mathrm{local}}$. We go on to describe this spectrum and wavefunctions in section \ref{sec:massivespec} and in doing so will rederive the formula (\ref{locchi}). 

Let us discuss the relation between global and local modes, as defined above, from the local patch perspective. It is important to note that although for global modes the chirality is not determined by the local value of the Higgs and flux, once we fix the chirality, say from some global model, then the local profile of the wavefunction is determined and so still information can be gained regarding operators involving such modes. Also we should note a calculational advantage of local modes over global modes which is that since the bulk of the wavefunction is located in the patch their normalisation can be determined more accurately. 

Since the flux does not determine the particular basis of holomorphic sections it can not be that on a local patch we can determine whether a mode is local or global by specifying only the flux and Higgs, but rather it must be an additional input parameter in the local model. Indeed we can see this quite generally from the general forms of the local Higgs (\ref{phi}) and flux (\ref{a}) by noting that under a coordinate transformation
\be
z_1 \rightarrow z_1 + m_2 a \;,\;\; z_2 \rightarrow z_2 - m_1 a \;, \label{locwiltrans}
\ee
for any complex constant $a$, the Higgs background remains invariant. The gauge field develops a constant term but this is pure gauge and so can be gauged away by a phase rotation of the wavefunction thereby leaving its peak unchanged. Hence locally it is not possible to determine the position of the wavefunction peak from the equations of motion since the peak can be arbitrarily displaced by the parameter $a$. The parameter $a$ is what we will call a local Wilson line, it is associated to the symmetry of the local flat metric on the patch, and if the metric remains flat globally so that the curve is a torus, then $a$ is promoted to a global Wilson line. From a local perspective $a$ is an input parameter that determines, amongst other things, whether a mode is local or global: if the peak of the wavefunction is displaced sufficiently away from $z_i=0$, which is where the interaction point is defined, then the mode is global, otherwise we denote it local. We will make these statements more precise in section \ref{sec:wave} where we also show that the parameter $a$ is part of the arbitrary holomorphic prefactor present in local solutions to the Dirac equation on flat space, as expected from the discussion above.

Finally, it is important to note that like other local parameters, the local Wilson line $a$ can be constrained by phenomenology. A direct way to see its effect is simply to note that moving the wavefunction peak away from the interaction point will decrease the coefficients of operators involving that mode. Indeed this is the local version of the mechanism proposed in \cite{Hayashi:2009bt} to generate flavour hierarchies. Such a local Wilson line may indeed be useful for suppressing operators but as expected there is a limit to how much this can be done since the size of $S_{GUT}$ is finite.\footnote{Note though that it may be possible to generate substantial suppression by deforming the local geometry, analogously to the findings of \cite{Hayashi:2009bt} where this was done with large complex-structure.} Within a local theory a conservative bound should be that the local Wilson line should not displace the wavefunction peak further than the boundary of the local patch. We will make some quantitative statements regarding such suppression in section \ref{sec:wave}.

\subsection{The spectrum}
\label{sec:massivespec}

Having discussed the relation between local and global wavefunctions and chirality in general terms, in this section we describe the concrete and explicit realisation of this relation and how it determines the form of the massless and massive wavefunctions.

The particular basis of the ${\cal N}=4$ fields given in the explicit form for $\Psi$ (\ref{z2}) do not correspond to four-dimensional mass eigenstates and therefore we first need to go to such an appropriate mass eigenstates basis. To do this we follow and expand on the procedure used in \cite{Marchesano:2010bs,Aparicio:2011jx,Camara:2011nj}. The Laplacian (\ref{masseom}) can be written as
\be
\mathbb{D}^+\mathbb{D}^- = -\Delta {\mathbb I} + {\mathbb B} \;, \label{laplacian}
\ee
where 
\be
\Delta = \sum_{i=1}^3 D_i^+ D_i^- \;,
\ee
and 
\be
\mathbb{B}=\begin{pmatrix}0&0&0&0\\
0&[D_2^+,D_2^-]&[D_2^-,D_1^+]&[D_3^-,D_1^+]\\
0&[D_1^-,D_2^+]&[D_1^+,D_1^-]&[D_3^-,D_2^+]\\
0&[D_1^-,D_3^+]&[D_2^-,D_3^+]&[D_2^+,D_2^-]+[D_1^+,D_1^-]
\end{pmatrix}\label{bmatrix} \;.
\ee
Using the D-term constraint it is simple to check that the Hermitian matrix ${\mathbb B}$ is also traceless. Let us denote the 3 normalised eigenvectors of the non-trivial block of ${\mathbb B}$ as $\hat{\xi}_p$ with associated real eigenvalues $\lambda_p$ which satisfy $\lambda_1 + \lambda_2+\lambda_3=0$. 
Diagonalising ${\mathbb B}$ with a unitary matrix ${\mathbb J}$ gives 
\be
{\mathbb J}^{\dagger}\cdot {\mathbb B}\cdot {\mathbb J} = \mathrm{diag}\left(0,\lambda_1,\lambda_2,\lambda_3\right)\;.
\ee
The associated rotation of the ${\mathbb D}^{\pm}$ matrices is
\be
\tilde{{\mathbb D}}^+ \equiv {\mathbb J}^{\dagger}\cdot{\mathbb D}^+\cdot{\mathbb J}^* \;,\;\;
\tilde{{\mathbb D}}^- \equiv {\mathbb J}^{T}\cdot{\mathbb D}^-\cdot{\mathbb J} \;.
\ee
The rotated matrices $\tilde{{\mathbb D}}^{\pm}$ take the same form as (\ref{z2}) but with $D_i^{\pm} \rightarrow \tilde{D}_i^{\pm}$ where
\be
\tilde{D}_p^{-} = \sum_j \hat{\xi}_{p,j} \; D_j^{-} \;,\;\; \tilde{D}_p^{+} = \sum_j \hat{\xi}^*_{p,j} \; D_j^{+} \;,
\ee
where $\hat{\xi}_{p,j}$ denotes the $j$th component of the $p$th eigenvector. In this basis the only non-trivial commutators are the diagonal ones
\be
\left[\tilde{D}_p^{+},\tilde{D}_q^{-} \right] = - \delta_{pq}\lambda_p \;. \label{commlam}
\ee
These commutation relations correspond to the algebra of 3 quantum harmonic oscillators. 

Let us now connect this to the general discussion of section \ref{sec:locspe}. The massless mode wavefunction in the diagonal basis satisfies the equation
\be
\mathbb{D}^- \Psi = \tilde{\mathbb{D}}^- \tilde{\Psi} = 0 \;,
\ee
where $\tilde{\Psi}=\mathbb{J}^{\dagger}\Psi$. Since we have diagonalised (\ref{laplacian}) the eigenvectors $\tilde{\Psi}$ are simply the vectors with one non-vanishing entry and so there are four of them. We denote them as
\be
\tilde{\Psi}_0 = \begin{pmatrix} \varphi_0 \\ 0\\ 0 \\ 0\end{pmatrix} \;, \;\; 
\tilde{\Psi}_1 = \begin{pmatrix} 0 \\ \varphi_1\\ 0 \\ 0\end{pmatrix} \;, \;\;
\tilde{\Psi}_2= \begin{pmatrix} 0\\ 0\\ \varphi_2 \\ 0\end{pmatrix} \;, \;\;
\tilde{\Psi}_3 = \begin{pmatrix} 0 \\ 0\\ 0 \\ \varphi_3 \end{pmatrix} \;.\label{psigen}
\ee
These are the 4 massless modes that form a full ${\cal N}=4$ multiplet in the absence of a Higgs background and flux. Each function $\varphi_P$, with $P=0,...,3$,  satisfies three equations (of which only 2 are independent) 
\bea
\tilde{D}^-_1 \varphi_0 &=& \tilde{D}^-_2 \varphi_0 = \tilde{D}^-_3 \varphi_0 = 0 \;,\nn \\
\tilde{D}^-_1 \varphi_1 &=& \tilde{D}^+_2 \varphi_1 = \tilde{D}^+_3 \varphi_1 = 0 \;, \nn \\
\tilde{D}^+_1 \varphi_2 &=& \tilde{D}^-_2 \varphi_2 = \tilde{D}^+_3 \varphi_2 = 0 \;, \nn \\
\tilde{D}^+_1 \varphi_3 &=& \tilde{D}^+_2 \varphi_3 = \tilde{D}^-_3 \varphi_3 = 0 \;.
\eea
As discussed in section \ref{sec:locspe} chirality arises upon requiring that the wavefunctions are normalisable. Consider first normalisability normal to the matter curve which is due to the Higgs profile. We define our local theory and patch so that the wavefunctions are strongly localised onto the curve and so we should see the this normalisability locally. It is simple to check, by turning off the flux, that $\varphi_0$ and $\varphi_3$ are projected out by this condition leaving an ${\cal N}=2$ multiplet worth of fields.

If we try to impose normalisability to deduce the chirality between $\varphi_1$ and $\varphi_2$ we can only do this for local modes as defined in section \ref{sec:locspe}. Global modes do not have a wavefunction turning point in the local patch and so there is no sense in which they are locally normalisable. For local modes the crucial objects for understanding the chirality are the eigenvalues $\lambda_i$. We go on to discuss these but first we note a constraint on the parameters which comes from the local geometric picture. In this paper we require that the curves are such that their width is much less than their length. In terms of the background parameters, where we take as is expected $m_i \sim M_{ij} \sim {\cal O}\left(1\right)$, this is the limit where $R \gg 1$. The geometric regime $R \sim {\cal O}(1)$ is only consistent with the constraints on the effective theory discussed in section \ref{sec:eqmot} if $R_{\perp} \ll 1$. Strictly speaking this is not necessarily a problem for the gauge theory as $R_{\perp}$ is a local scale while gravitational corrections, which are the key problem in this regime, are expected to be controlled by more global scales. Nonetheless it seems difficult to realise such a setup within a well controlled geometric regime and so we do not study this region of parameter space in much detail henceforth.

In the limit $R \gg 1$ two of the eigenvalues, $\lambda_1$ and $\lambda_2$ scale as $R$ and so become much larger than $\lambda_3$ which does not scale with $R$. Since their sum vanishes it implies $\lambda_1 \simeq -\lambda_2$ and we take by definition $\lambda_1<0$. 
The small difference between the eigenvalues $\lambda_1$ and $\lambda_2$, or equivalently the values of $\lambda_3$ is set by the background flux. We therefore can identify 3 possibilities on their sign distribution according to the sign of the determinant of the matrix ${\mathbb B}$
\bea
\mathrm{det}\;{\mathbb B} < 0 &\implies & \lambda_1 < 0 \;,\;\; \lambda_2 > 0 \;,\;\; \lambda_3 > 0 \;, \nn \\
\mathrm{det}\;{\mathbb B} > 0 &\implies & \lambda_1 < 0 \;,\;\; \lambda_2 > 0 \;,\;\; \lambda_3 < 0 \;, \nn \\
\mathrm{det}\;{\mathbb B} = 0 &\implies & \lambda_1 < 0 \;,\;\; \lambda_2 > 0 \;,\;\; \lambda_3 = 0 \;.
\eea
We discuss the nature of the wavefunctions for these 3 possibilities below. 

\subsubsection{$\mathrm{det}\;{\mathbb B} \neq 0$\;: localised ${\cal N}=1$ spectrum}
\label{sec:n1spec}

Consider the case $\mathrm{det}\;{\mathbb B} < 0$. In this case chirality is generated because $\varphi_2$ is projected out by the local criteria of normalisability. It is possible to see this by studying the wavefunction explicitly but there is neater approach which will also be useful for our further discussion. Note that in this case we see that the commutation relations (\ref{commlam}) imply that we should identify the raising and lowering operators of the harmonic oscillators, denoted $\hat{a}_i^+$ and $\hat{a}_i^-$ respectively, as
\begin{align}
\hat{a}^-_1 &\equiv i\tilde D_1^-\;, &
\hat{a}^-_2 &\equiv i\tilde D_2^+ \;, &
\hat{a}^-_3 &\equiv i\tilde D_3^+ \;, \nn \\
\hat{a}^+_1 &\equiv i\tilde D^+_1\;,  &
\hat{a}^+_2 &\equiv i\tilde D_2^-\;, &
\hat{a}^+_3 &\equiv i\tilde D_3^-\;. \label{dertoop}
\end{align}
With these identifications we can write the Hamiltonian as
\be
\tilde{\mathbb{D}}^+\tilde{\mathbb{D}}^- = \sum_{i=1}^3 {\hat{a}_i^+ \hat{a}_i^-} \;\mathbb{I}+\textrm{diag}(-\lambda_1,0,\lambda_2-\lambda_1,\lambda_3-\lambda_1)\label{ddtilde} \;.
\ee
The components of $\Psi$ form 4 towers of modes and for each tower the lowering operators $\hat{a}^-_i$ annihilate the ground state and so the last term of (\ref{ddtilde}) gives the masses of the ground states. We see that this picks out the $\varphi_1$ function and so there is just one massless ground state corresponding to a single ${\cal N}=1$ chiral multiplet, i.e. the spectrum is chiral. The particular one which is annihilated by the operators (\ref{dertoop}) is
\be
\tilde{\mathbb{D}}^-\cdot\begin{pmatrix}0\\ \varphi\\ 0\\ 0\end{pmatrix}= 0\quad \Leftrightarrow \quad \begin{cases}\tilde D^-_1\varphi&=0\\
\tilde{D}_2^+\varphi&=0\\
\tilde{D}_3^+\varphi&=0\end{cases} \;.\label{zero00}  
\ee
These are the equations that we solve in section \ref{sec:wave} to determine the form of the wavefunction. The other three eigenvectors of the diagonalised version of the Laplacian (\ref{laplacian}) are of the same form as (\ref{zero00}) but with $\varphi$ being along the other three entries.
Returning to the original non-diagonal basis therefore gives for the ground states 
\be
\Psi_0 = \frac{1}{N} \begin{pmatrix} 1 \\ 0\\ 0 \\ 0\end{pmatrix} \; \varphi \;, \;\; \Psi_p = \frac{1}{N} \begin{pmatrix} 0 \\ \hat{\xi}_{p,1} \\ \hat{\xi}_{p,2} \\ \hat{\xi}_{p,3}\end{pmatrix} \; \varphi \;, \label{psiphi}
\ee
where $N$ is an appropriate normalisation factor. 
The massive excited states of the towers, the Landau-levels, are simply obtained by applying the raising operators which gives
\be
\Psi_{P,(n,m,l)}=\frac{\left(i\tilde{D}_1^+\right)^{n}\left(i\tilde D_2^-\right)^m\left(i\tilde D_3^-\right)^l }{\sqrt{m!n!l!}\left(-\lambda_1\right)^{n/2}\lambda_2^{m/2}\lambda_3^{l/2}}\Psi_{P}\;,\label{massiverep}
\ee
with $P=0,...,3$.
The excited states have masses given by\footnote{Note that the mass spectrum is not quite of a quantum harmonic oscillator since the ground state is massless. This shift in the spectrum is another way to see the requirement of the twisting of the 8-dimensional theory for supersymmetry.}
\begin{align}
M^2_{\Psi_{0,(n,m,l)}} &= M_K^2 \left[ -(n+1)\lambda_1+m\lambda_2+l\lambda_3 \right] \;, \label{mas}\\
M^2_{\Psi_{1,(n,m,l)}} &= M_K^2 \left( -n\lambda_1+m\lambda_2+l\lambda_3 \right) \;, \nn\\
M^2_{\Psi_{2,(n,m,l)}} &= M_K^2 \left[ -(n+1)\lambda_1+(m+1)\lambda_2+l\lambda_3 \right] \;, \nn\\
M^2_{\Psi_{3,(n,m,l)}} &= M_K^2 \left[ -(n+1)\lambda_1+m\lambda_2+(l+1)\lambda_3 \right] \;. \nn
\end{align}
It is useful to see explicitly the mass operator in the theory since the pairing is non-trivial. The mass which appears in the 4-dimensional superpotential takes the form
\be
W^M_{ab} = i\int_S \mathrm{Tr\;} \left[ \left(\Psi_a^{\bf \bar{R}}\right)^T \mathbb{D}^- \Psi_b^{\bf R} \right] = i\int_S \mathrm{Tr\;} \left[ \left(\tilde{\Psi}_a^{\bf \bar{R}}\right)^T \tilde{\mathbb{D}}^- \tilde{\Psi}_b^{\bf R} \right] \;, \label{massterm}
\ee
where the indices $a,b$ label 4-dimensional fields and here $\Psi$ are their internal wavefunction profiles.
The mass pairing is between a representation ${\bf R}$ and its conjugate ${\bf \bar{R}}$. Note that by ${\bf \bar{R}}$ we refer to the ${\cal N}=2$ partner of the state rather than the anti-particle which can not appear in a superpotential mass. The wavefunctions for the ${\bf \bar{R}}$ states are the complex conjugates of the ${\bf R}$ states described above
\be
\Psi_{P,\left(n,m,l\right)}^{\bf \bar{R}} = \left(\Psi_{P,\left(n,m,l\right)}^{\bf R}\right)^* \;, \label{rrbcc} 
\ee
but since the charge is flipped the annihilation and creation operators are interchanged so that now the first term on the right-hand-side of (\ref{ddtilde}) does not annihilate the state and so after performing the normal ordering there is a mass shift in the states so that 
\bea
M^2_{\bf R} &=& M_K^2\left\{-\lambda_1,0,\lambda_2-\lambda_1,\lambda_3-\lambda_1\right\} \;, \nn \\
M^2_{{\bf \bar{R}}} &=& M_K^2\left\{-\lambda_1,-2\lambda_1,\lambda_3,\lambda_2\right\} \;. \label{rrbmass}
\eea
These are the ground states of the ${\bf \bar{R}}$ modes, the excited states are created with the annihilation operators of the ${\bf R}$ states (\ref{zero00}) and have increasing masses as in (\ref{mas}). Note that we will generally drop the ${\bf R}$ and ${\bf \bar{R}}$ superscripts unless explicitly required with the default state being ${\bf R}$.

The operator $\tilde{\mathbb{D}}^-$ in the mass term (\ref{massterm}) acts as a raising/lowering operator and so the coupling is between states of different levels and also of different ${\cal N}=4$ towers. Since this plays an important role in generating non-renormalisable operators in the theory after integrating out massive modes it is worth discussing it in more detail. The relation used to determine the pairing and the mass is
\bea
i\mathbb{D}^- \Psi^{\bf R}_{0,(n,m,l)} &=& -\sqrt{n} \left(-\lambda_1\right)^{\frac12} \Psi^{\bf R}_{1,(n-1,m,l)} - \sqrt{m+1} \left(\lambda_2\right)^{\frac12} \Psi^{\bf R}_{2,(n,m+1,l)}  - \sqrt{l+1} \left(\lambda_3\right)^{\frac12} \Psi^{\bf R}_{3,(n,m,l+1)} \;, \nn \\
i\mathbb{D}^- \Psi^{\bf R}_{1,(n,m,l)} &=& \sqrt{n} \left(-\lambda_1\right)^{\frac12} \Psi^{\bf R}_{0,(n-1,m,l)} - \sqrt{m} \left(\lambda_2\right)^{\frac12} \Psi^{\bf R}_{3,(n,m-1,l)}  + \sqrt{l} \left(\lambda_3\right)^{\frac12} \Psi^{\bf R}_{2,(n,m,l-1)} \;,  \label{dpsi} \\
i\mathbb{D}^- \Psi^{\bf R}_{2,(n,m,l)} &=& \sqrt{n+1} \left(-\lambda_1\right)^{\frac12} \Psi^{\bf R}_{3,(n+1,m,l)} + \sqrt{m+1} \left(\lambda_2\right)^{\frac12} \Psi^{\bf R}_{0,(n,m+1,l)}  - \sqrt{l} \left(\lambda_3\right)^{\frac12} \Psi^{\bf R}_{1,(n,m,l-1)} \;, \nn \\
i\mathbb{D}^- \Psi^{\bf R}_{3,(n,m,l)} &=& -\sqrt{n+1} \left(-\lambda_1\right)^{\frac12} \Psi^{\bf R}_{2,(n+1,m,l)} + \sqrt{m} \left(\lambda_2\right)^{\frac12} \Psi^{\bf R}_{1,(n,m-1,l)}  + \sqrt{l+1} \left(\lambda_3\right)^{\frac12} \Psi^{\bf R}_{0,(n,m,l+1)} \;. \nn
\eea
The massive modes form mass degenerate groups within which the states couple to each other. Let us discuss one such group as an example. First we make the external four-dimensional field $\phi$ and its internal wavefunction $\Psi$ manifest by using the notation 
\be
\hat{\Psi}^{\bf R}_{P,(n,m,l)} = \phi^{\bf R}_{P,(n,m,l)}\left(x_{\mu}\right) \Psi^{\bf R}_{P,(n,m,l)}\left(z_i\right)  \;.
\ee
Then using (\ref{dpsi}) and (\ref{massterm}) we find the following four-dimensional superpotential interaction\footnote{Note that here we refer to the `physical' rather than holomorphic superpotential in the sense that the fields are canonically normalised.}
\be
W \supset i \left(\Phi_0^{\bf R} \right)^T \cdot {\bf M} \cdot \Phi_0^{\bf R} \;,
\ee
where
\bea
\Phi_0^{\bf R} &=& \begin{pmatrix} \phi^{\bf R}_{0,(0,0,0)} \\ \phi^{\bf \bar{R}}_{2,(0,1,0)} \\ \phi^{\bf \bar{R}}_{3,(0,0,1)} \\ \phi^{\bf R}_{1,(0,1,1)}\end{pmatrix} \;, \;\;{\bf M} = M_K\begin{pmatrix} 
0 & -\left(\lambda_2\right)^{\frac12} & -\left(\lambda_3\right)^{\frac12} & 0 \\
-\left(\lambda_2\right)^{\frac12} & 0 & 0 & \left(\lambda_3\right)^{\frac12} \\ 
-\left(\lambda_3\right)^{\frac12} & 0 & 0 & -\left(\lambda_2\right)^{\frac12} \\ 
0 & \left(\lambda_3\right)^{\frac12} & -\left(\lambda_2\right)^{\frac12} & 0 \end{pmatrix} \;. \label{pmexp}
\eea
The coupling gives the appropriate degenerate mass matrix
\be
{\bf M}^2 = M_K^2 \left(\lambda_2+\lambda_3\right) \mathbf{I} \;,
\ee
which indeed matches the masses calculated using (\ref{rrbmass}). Similar groupings occur for the other massive modes, and the pattern is replicated at each level of the tower. Explicitly for some level in the tower labeled by $n \geq 0$ we have the groupings 
\bea
\Phi_2^{\bf R} &=& \begin{pmatrix} \phi^{\bf R}_{2,(n,n,n)} \\ \phi^{\bf \bar{R}}_{3,(n+1,n,n)} \\ \phi^{\bf \bar{R}}_{0,(n,n+1,n)} \\ \phi^{\bf R}_{1,(n+1,n+1,n)}\end{pmatrix} \;, \;\;
\Phi_3^{\bf R} = \begin{pmatrix} \phi^{\bf R}_{3,(n,n,n)} \\ \phi^{\bf \bar{R}}_{2,(n+1,n,n)} \\ \phi^{\bf \bar{R}}_{0,(n,n,n+1)} \\ \phi^{\bf R}_{1,(n+1,n,n+1)}\end{pmatrix} \;, \;\;
\Phi_1^{\bf \bar{R}} = \begin{pmatrix} \phi^{\bf \bar{R}}_{1,(n,n,n)} \\ \phi^{\bf R}_{3,(n,n+1,n)} \\ \phi^{\bf R}_{2,(n,n,n+1)} \\ \phi^{\bf \bar{R}}_{0,(n,n+1,n+1)}\end{pmatrix} \;, \nn \\ 
\Phi_0^{\bf \bar{R}} &=& \begin{pmatrix} \phi^{\bf \bar{R}}_{0,(n,n,n)} \\ \phi^{\bf R}_{1,(n+1,n,n)} \end{pmatrix} \;, \;\;\;\;\;\;\;
\Phi_2^{\bf \bar{R}} = \begin{pmatrix} \phi^{\bf \bar{R}}_{2,(n,n,n)} \\ \phi^{\bf R}_{1,(n,n,n+1)} \end{pmatrix} \;,  \;\;\;\;\;\;\;
\Phi_3^{\bf \bar{R}} = \begin{pmatrix} \phi^{\bf \bar{R}}_{3,(n,n,n)} \\ \phi^{\bf R}_{1,(n,n+1,n)} \end{pmatrix} \;.  \label{llgrps}
\eea
The pattern of coupling within the groups is the same as in (\ref{pmexp}): each state couples to the state with one more or less excitation. In integrating out massive modes therefore we must make sure that they appropriately couple to each other as in the grouping above.

The opposite case with $\mathrm{det}\;{\mathbb B} > 0$ is very similar except that now the massless localised mode is in the conjugate representation. So if we fix the charges $q^a$ the localised mode would be the one with $-q^a$. To summarise we showed that the sign of $\mathrm{det}\;{\mathbb B}$ determines the chirality of a localised ${\cal N}=1$ mode. This then directly leads to a rederivation of the formula (\ref{locchi}) where we define
\be
\chi_{\mathrm{local}} = -\frac{\mathrm{det}\;{\mathbb B}}{R^2}  \;.
\ee

\subsubsection{$\mathrm{det}\;{\mathbb B} = 0$\;: delocalised ${\cal N}=2$ spectrum}

In this case the spectrum changes structure. Since $\lambda_3=0$ we see that $\tilde{D}_3^{\pm}$ are no longer raising and lowering operators but rather commute with all the operators and so correspond to a conserved charge. The new conserved quantum number labels the Kaluza-Klein (KK) momentum along the curve. Now all the modes, including the massless ones, form ${\cal N}=2$ representations and the spectrum is non-chiral. 

To see this we can compare the masses of the ground state (\ref{rrbmass}) which show that as $\lambda_3 \rightarrow 0$ the spectrum becomes vector-like at the massless and massive levels. We now have 2 massless modes
\be
{\mathbb D}^- \Psi_1^{\bf R} = {\mathbb D}^- \Psi_2^{\bf \bar{R}} = 0 \;.
\ee
These are the massless modes corresponding to vanishing KK momentum. There is also a KK tower which generalises the equations above to 
\bea
i{\mathbb D}^- \Psi_1^{\bf R} = k_K \Psi_2^{\bf R}  \;, \;\;\;
i{\mathbb D}^- \Psi_2^{\bf \bar{R}} = k_K \Psi_1^{\bf \bar{R}}  \;. \label{kkrm}
\eea
The wavefunctions still take the form (\ref{psiphi}) but now the equations for $\varphi$ read
\be
\tilde{D}_1^- \varphi = 0 \;,\;\; \tilde{D}_2^+ \varphi = 0 \;,\;\; \tilde{D}_3^{\pm} \varphi = -i k_K^{\pm} \varphi \;. \label{kkphieq}
\ee
Now $k_K^{\pm}$ are conserved quantum numbers associated to the KK number, but note that in the local theory they are not quantised. We can define $k_K= k_K^+ = \left(k_K^-\right)^*$ and the associated mass operator is
\be
\tilde{D}_3^+\tilde{D}_3^- \varphi = -|k_K|^2 \varphi \;,
\ee
with physical mass $M_K \left|k_K\right|$.
We can still act with the two Landau-level raising operators to create excited modes as before so that the modes takes the form
\begin{equation}
\Psi_{P,(n,m,k_K)}=\frac{(i\tilde{D}_1^+)^{n}(i\tilde D_2^-)^m}{\sqrt{m!n!}(-\lambda_1)^{n/2}\left(\lambda_2\right)^{m/2}}\Psi_{P,(0,0,k_K)} \;,
\end{equation}
with associated masses
\begin{align}
M^2_{\Psi_{0,(n,m,k_K)}} &= M_K^2 \left[ -(m+n+1)\lambda_1+|k_K|^2 \right] \;, \\
M^2_{\Psi_{1,(n,m,k_K)}} &= M_K^2 \left[ -(m+n)\lambda_1+|k_K|^2 \right] \;, \nn\\
M^2_{\Psi_{2,(n,m,k_K)}} &= M_K^2 \left[ -(m+n+2)\lambda_1+|k_K|^2 \right] \;, \nn\\
M^2_{\Psi_{3,(n,m,k_K)}} &= M_K^2 \left[ -(m+n+1)\lambda_1+|k_K|^2 \right] \;. \nn
\end{align}
There is a similar grouping of massive modes as in (\ref{pmexp}) which can be deduced using (\ref{kkrm}). For vanishing LL excitations, which are the only cases we will use, the pairing are
\be
\Phi_0^{\bf R} = \begin{pmatrix} \phi^{\bf R}_{0,(0,0,k_K)} \\ \phi^{\bf \bar{R}}_{3,(0,0,k_K)} \end{pmatrix} \;, 
\Phi_1^{\bf R} = \begin{pmatrix} \phi^{\bf R}_{1,(0,0,k_K)} \\ \phi^{\bf \bar{R}}_{2,(0,0,k_K)} \end{pmatrix} \;,
\Phi_2^{\bf R} = \begin{pmatrix} \phi^{\bf R}_{2,(0,0,k_K)} \\ \phi^{\bf \bar{R}}_{1,(0,0,k_K)} \end{pmatrix} \;,
\Phi_3^{\bf R} = \begin{pmatrix} \phi^{\bf R}_{3,(0,0,k_K)} \\ \phi^{\bf \bar{R}}_{0,(0,0,k_K)} \end{pmatrix} \;. \label{kkpairings}
\ee

The wavefunctions develop an isometry, and therefore delocalise, in this case which can be seen by noting that we can write 
\be
\tilde{D}_3^{-} = \left(\hx_{3,1}\partial_{z_1}+ \hx_{3,2}\partial_{z_2}\right) - \bar{g} \;,\;\;\tilde{D}_3^{+} = \left(\hx_{3,1}^*\bar{\partial}_{\bar{z}_1}+ \hx_{3,2}^*\bar{\partial}_{\bar{z}_2}\right) + g \;,
\ee
where $g$ is some function of the flux and Higgs vevs. Then using (\ref{kkphieq}) we have that
\bea
\left(\hx_{3,1}\partial_{z_1}+ \hx_{3,2}\partial_{z_2}\right) \left|\vphi\right|^2 &=&  \bar{\vphi} \left(\hx_{3,1}\partial_{z_1}+ \hx_{3,2}\partial_{z_2}\right) \vphi + \vphi \left[\left(\hx_{3,1}^*\bar{\partial}_{\bar{z}_1}+ \hx_{3,2}^*\bar{\partial}_{\bar{z}_2}\right) \vphi\right]^* \nn \\ 
&=&  \left|\vphi\right|^2 \left[ \left(-i\bar{k}_K + \bar{g}\right) + \left( i\bar{k}_K-\bar{g}\right) \right] \nn \\
&=& 0 \;. \label{kkisogen}
\eea
This delocalisation has of course implications for determining the normalisation of the wavefunctions as discussed in section \ref{sec:wavekk}. Also the delocalisation shows that these modes are not local in any sense, indeed the presence of the massless vector pair of zero modes can not be determined locally and depends on the boundary conditions. 

The delocalisation also has important implications for determining the chiral nature of the spectrum. Although for the ${\cal N}=1$ modes the chirality is determined by evaluating $\chi_{\mathrm{local}}$ at the wavefunction maximum, for the ${\cal N}=2$ modes, because of the isometry, we require that $\chi_{\mathrm{local}}=0$ throughout the full matter curve and in particular also at the point of $E_8$.

\section{Wavefunctions and operators}
\label{sec:wave}

In the previous section we set up the local effective theory and showed that some information regarding the local massless spectrum can be obtained. Moreover the primary advantage of the local theory is the possibility to calculate coefficients of operators in the four-dimensional theory. Such coefficients come from overlap integrals of internal wavefunctions. In this section we calculate the wavefunctions and their overlaps as a function of general Higgs vev and fluxes.

\subsection{Landau-level wavefunctions}
\label{sec:wavell} 

Landau-level wavefunctions correspond to the case where $\chi_{\mathrm{local}}\neq0$ and by convention we choose it to be positive so that, if the massless mode is local, in the sense defined in section \ref{sec:locspe}, then we know the appropriate wavefunction solves the equations (\ref{zero00}). If the mode is global then the local flux does not determine the chirality of the mode, and therefore does not uniquely determine the wavefunction. However for concreteness we will still take the wavefunction to be given by $\varphi_1$ in (\ref{psigen}) and will include the effect of local Wilson lines. It may be that globally the massless wavefunction is actually $\varphi_2$ in (\ref{psigen}) and this would correspond to the same form of wavefunction but with the charge of the mode flipped.  

In order to determine the wavefunctions for the fields as in (\ref{psiphi}) we need to solve for the vectors $\hat{\xi}_p$ and for the wavefunction profile $\varphi$. The former are eigenvectors of the matrix $\mathbb{B}$ which reads
\be
\mathbb{B}=M_{K}^2\begin{pmatrix}0&0&0&0\\
0&\half \left(\bar{M}_{11} + M_{11}\right) & \half \left(\bar{M}_{12} + M_{21}\right) & R\bar{m}_1 \\
0& \half \left(\bar{M}_{21} + M_{12}\right)  &\half \left(\bar{M}_{22} + M_{22}\right)& R \bar{m}_2\\
0&Rm_1&Rm_2& 0 \end{pmatrix}\label{bmatrix2} \;.
\ee
The non-trivial eigenvalues of the matrix $\lambda_i$ satisfy the cubic equation
\bea
-\lambda_i^3 &+& \lambda_i \left(R^2|m_1|^2 + R^2|m_2|^2 + \frac14 \left|M_{12}+\bar{M}_{21} \right|^2 +\frac14\left|M_{11}+\bar{M}_{11} \right|^2  \right) \nn \\ & & + R^2 \;\mathrm{Re}\left[\left(M_{12}+\bar{M}_{21}\right)\bar{m}_1 m_2 + M_{11} \left(\left|m_1\right|^2-\left|m_2\right|^2\right)\right] =0 \;.
\eea
Although readily obtainable, the solutions to this equation take a rather complicated form and so in the following we calculate in terms of the implicit solutions for $\lambda_i$. A compact expression for the eigenvectors in terms of the eigenvalues is 
\be
\xi_i=\begin{pmatrix} \half \left( \bar{M}_{12}+M_{21} \right)\lambda_i + R^2 m_2 \bar{m}_1 \\ 
\lambda_i^2 - \half \left(\bar{M}_{11}+M_{11} \right)\lambda_i - R^2 |m_1|^2 \\ 
R m_2 \lambda_i+ \half R m_1 \left( \bar{M}_{12}+M_{21} \right) - \half R m_2  \left(\bar{M}_{11}+M_{11} \right)  \end{pmatrix} \;, \label{eivecb}
\ee
and we define $\hat{\xi}_i=\xi_i/\left|\xi_i\right|$. There is a subtlety due to the fact that for non-generic choices of fluxes (for example $M_{12}=m_2=0$) some of the vectors may degenerate in which case the dual expressions should be used
\be
\xi_i=\begin{pmatrix} \lambda_i^2 + \half \left(\bar{M}_{11}+M_{11} \right)\lambda_i - R^2 |m_2|^2 \\ 
\half \left( \bar{M}_{21}+M_{12} \right)\lambda_i + R^2 m_1 \bar{m}_2 \\ 
R m_1 \lambda_i+ \half R m_2 \left( \bar{M}_{21}+M_{12} \right) + \half R m_1  \left(\bar{M}_{11}+M_{11} \right)  \end{pmatrix} \;. \label{eivecb2}
\ee
The wavefunction profile is obtained by solving (\ref{zero00}) and takes the form
\be
\vphi = f\left(-\hx_{1,2}z_1 + \hx_{1,1} z_2\right)e^{-p_1 |z_1|^2 - p_2 |z_2|^2 + p_3 \bar{z}_1z_2 + p_4 \bar{z}_2 z_1} \;,\label{vphi}
\ee
where we have
\bea
p_1 &=& \half M_{11} - R m_1 \left(\frac{\hx^*_{3,2} \hx^*_{2,3} - \hx^*_{2,2} \hx^*_{3,3}}{\hx^*_{3,1} \hx^*_{2,2} - \hx^*_{2,1} \hx^*_{3,2}}\right) \;, \nn \\
p_2 &=& \half M_{22} + R m_2 \left(\frac{\hx^*_{3,1} \hx^*_{2,3} - \hx^*_{2,1} \hx^*_{3,3}}{\hx^*_{3,1} \hx^*_{2,2} - \hx^*_{2,1} \hx^*_{3,2}}\right) \;, \nn \\
p_3 &=& -\half M_{21} + R m_2 \left(\frac{\hx^*_{3,2} \hx^*_{2,3} - \hx^*_{2,2} \hx^*_{3,3}}{\hx^*_{3,1} \hx^*_{2,2} - \hx^*_{2,1} \hx^*_{3,2}}\right) \;, \nn \\
p_4 &=& -\half M_{12} - R m_1 \left(\frac{\hx^*_{3,1} \hx^*_{2,3} - \hx^*_{2,1} \hx^*_{3,3}}{\hx^*_{3,1} \hx^*_{2,2} - \hx^*_{2,1} \hx^*_{3,2}}\right) \;.
\eea
It is worth noting that the pure flux factors in the above expressions all cancel out in a charge neutral triple coupling because they appear linearly. Because of this they only affect the normalisation of the wavefunctions.

\subsubsection*{Holomorphic prefactor and local Wilson lines}

As discussed in section \ref{sec:locchir}, there is a local symmetry of the equations of motion which implies the presence of local freedom in the wavefunctions which we termed local Wilson lines. The values of these parameters determines the position of the wavefunction peak and so is the local parameter that differentiates between local and global modes as defined in section \ref{sec:locspe}. We can determine how the parameter $a$ defined in (\ref{locwiltrans}) must appear in the wavefunction. Starting from (\ref{vphi}), performing the coordinate transformation (\ref{locwiltrans}), and then performing a gauge transformation so that the background Higgs and flux take their original values we find that the wavefunction transforms as
\be
\varphi \rightarrow  \varphi\; C_{wl}\left(a\right)\varphi_{wl}\left(a\right) \;,
\ee
where we define
\bea
C_{wl} &=& e^{-|a|^2\left(p_1 |m_2|^2 + p_2 |m_1|^2 + p_3 m_1 \bar{m}_2 + p_4 m_2 \bar{m}_1  \right)}\;, \nn \\
\varphi_{wl} &=& e^{z_1 c_1 + z_2 c_2} \;, \label{cwlpf}
\eea
and
\bea
c_1 &=& \bar{a} \left[ \left(-p_1 - \frac12 \bar{M}_{11}\right)\bar{m}_2 +  \left(-p_4 + \frac12 \bar{M}_{21}\right)\bar{m}_1 \right] \;, \nn \\
c_2 &=& \bar{a} \left[ \left(p_2 + \frac12 \bar{M}_{22}\right)\bar{m}_1 +  \left(p_3 - \frac12 \bar{M}_{12}\right)\bar{m}_2 \right] \;.
\eea
It can be checked that the holomorphic function $\varphi_{wl}$ is a function of the particular combination of $z_1$ and $z_2$ that appears in the general holomorphic prefactor in (\ref{vphi}). From the transformation (\ref{locwiltrans}) we know that when $a \sim 1/m_1 \sim 1/m_2$ the wavefunction peak is displaced outside of the local patch and the mode should then be labeled as global. Since displacing the wavefunction peak along the matter curve generically is not expected to change normalisation of the wavefunction too much, the prefactor $C_{wl}$ gives an estimate of how much suppression of the wavefunction can be achieved by displacing it in the local patch, with an upper limit on $a$ as above. Given explicit values for the background Higgs and flux, this maximum possible suppression can be quantitatively analysed.

The holomorphic prefactor is directly related to the generation structure induced by the flux. Indeed we can account for this by putting a generation index on the local Wilson line parameter $a_l$. The actually global form of $f_l\left(-\hx_{1,2}z_1 + \hx_{1,1} z_2\right)$ will vary depending on the global topology of the matter curve. In the example studied in section \ref{sec:locspe} it took the form of a degree $l\leq M$ polynomial, where $M$ is the integrated value of the flux. 
We have some choice as to how we parameterise the generation dependence of the holomorphic prefactor in a local patch. It is possible to work directly with $\varphi_{wl}(a_l)$ which gives an explicit form for the peak of the wavefunction for each generation. 

A different basis within the generation space, is one where we take linear combinations of polynomials to successively eliminate all the lower powers of $\left(-\hx_{1,2}z_1 + \hx_{1,1} z_2\right)$ so that the polynomial representative for each generation has a single power equivalent to the generation number
\be
f_l\left(-\hx_{1,2}z_1 + \hx_{1,1} z_2\right) = \left(-\hx_{1,2}z_1 + \hx_{1,1} z_2\right)^l \;. \label{polyexp}
\ee
This choice of assigning the polynomial degree to the generation index was advocated in \cite{Heckman:2008qa} and subsequent studies \cite{Font:2009gq,Cecotti:2009zf,Aparicio:2011jx,Camara:2011nj} showed that it is a relevant parameterisation for some models of flavour such that the higher polynomials imply a smaller Yukawa coupling (and smaller coupling to massive modes). More generally the form (\ref{polyexp}) may also have higher order terms, this is the case on a torus where the holomorphic basis is formed by exponentials \cite{Cremades:2004wa} that are essentially the same form as the local Wilson line factors. Whether higher order terms are present or not in (\ref{polyexp}) can play an important role in some couplings as studied in section \ref{sec:waveover}. Note that since these higher powers only give subdominant contributions to the normalisation integral, and so essentially decouple from the normalisation, this leads to the interesting possibility of suppressing couplings by suppressing the higher order powers.

In presenting expressions we will allow for both approaches of parameterising the generation structure. So we will consider a holomorphic prefactor of the form 
\be
f_l\left(-\hx_{1,2}z_1 + \hx_{1,1} z_2\right) = \left(-\hx_{1,2}z_1 + \hx_{1,1} z_2\right)^l C_{wl}\left(a\right) \varphi_{wl}\left(a\right) \;. \label{genprehol}
\ee
According to the preferred basis then different generations can be studied either by setting $a=0$ and using the polynomial basis, or setting the generation index $l=0$ but then accounting for different values for $a$ for different generations. 

Note that, like the normalisation, the orthogonality of wavefunctions within generation space is a global question. One way to approach this issue is to assume global orthogonality by definition and thereby fix the local Wilson lines which will generically be non-vanishing for all the generations. The approach of taking linear combinations of wavefunctions so that the leading power denotes the generation number is incompatible with the assumption of global orthogonality in general. It may be that in some cases the two do match, such as the ${\mathbb P}^1$ example of section \ref{sec:globex}, but in general the higher order power in $f_l$ will induce mixing. However we can expect this mixing to be suppressed since the mixing integrals will involve higher powers of the $z_i$ compared to the diagonal normalisation integrals. 

\subsubsection*{Landau-Level wavefunction normalisation}
  
The remaining components of $\Psi_P$ as defined in (\ref{psiphi}) which are required are the normalisation factors. Since the vectors in (\ref{psiphi}) are normalised the normalisation factor simply comes from integrating $\left|\varphi\right|^2$ over $S$. There are two components to this integral: the integration orthogonal to the matter curve and the integration along the matter curve. In the regime we are working in, where $R \gg 1$ the localisation of the wavefunction perpendicular to the matter curve is strong which means we can perform that integral as an integral up to infinity over flat space, this being the local approximation of $S$, and expect that this does not receive significant corrections from the finite size of $S$. Regarding the integration along the matter curve, as discussed above, we present the result as an integral over infinity. This is expected to be valid for small Wilson lines and for substantial flux localisation where most of the integral contribution comes from the local patch. We should keep in mind that the integral result should not exceed the case where the wavefunction is completely delocalised along the matter curve, which in the homogeneous approximation is given approximately by (\ref{normkknofl}). This latter expression may be more accurate in the case of large Wilson lines or weak flux localisation. Note however that, quite generally, since the localisation within the curve by the flux is typically rather mild the normalisation integral for a homogeneous wavefunction is usually not a bad approximation. Indeed the difference between the normalisation integral as calculated with flux localisation and without localisation was studied in appendix B of \cite{Camara:2011nj} and was found to be relatively mild.

In general all the wavefunction overlap integrals will be expressed in terms of a Gaussian integral which we define as
\bea
& &I\left(n_1,n_2,n_3,n_4;p_1,p_2,p_3,p_4;a_1,a_2,b_1,b_2\right)\label{idef} \\
& &\equiv \int_{S} z_1^{n_1}\bar{z}_1^{n_2}z_2^{n_3}\bar{z}_2^{n_4}e^{-p_1|z_1|^2-p_2|z_2|^2+p_3z_2\bar{z}_1+p_4\bar{z}_2z_1 + a_1 z_1 + a_2 z_2 + b_1 \bar{z}_1 + b_2 \bar{z}_2 } \nn \\
& &= R^4_{\parallel}\left(\frac{\partial}{\partial a_1}\right)^{n_1} \left(\frac{\partial}{\partial b_1}\right)^{n_2} \left(\frac{\partial}{\partial a_2}\right)^{n_3} \left(\frac{\partial}{\partial b_2}\right)^{n_4}
\left[\frac{\pi^2}{p_1p_2-p_3p_4}
e^{\frac{a_2b_2p_1 + a_1b_1p_2+a_1b_2p_3+a_2b_1p_4}{p_1p_2-p_3p_4}} \right] \;, \nn
\eea
where the evaluation is done in the homogeneous flat-space non-compact approximation for $S$. It is important to note that the result of the integral is only correct if it satisfies the appropriate convergence properties. 

The normalisation of the wavefunction for generation index $l$ with prefactor (\ref{genprehol}) is then given by
\bea
N_l^2 &=& \left|C_{wl}\left(a\right) \right|^2\sum_{k_i} \frac{l!}{\prod_i \left(k_i!\right)} \left|\hx_{1,2}\right|^{2k_1}\left|\hx_{1,1}\right|^{2k_2}\left(-\hx_{1,2}\hx^*_{1,1}\right)^{k_3}\left(-\hx_{1,1}\hx^*_{1,2}\right)^{k_4}  \nn \\
& & \times I\left(k_1+k_3,k_1+k_4,k_2+k_4,k_2+k_3;p_1+\bar{p}_1,p_2+\bar{p}_2,p_3+\bar{p}_4,p_4+\bar{p}_3; \right.\nn \\
& & \;\;\;\;\;\;\; \left.c_1,c_2,\bar{c}_1,\bar{c}_2\right) \;, \nn \label{normland}  
\eea
where the $k_i$ are positive integers, with $i=1,...,4$, which are summed over all combinations satisfying the constraint $\sum^{i=4}_{i=1} k_i = l$.
Again we note that the result for the normalisation is only correct for convergent wavefunctions, while for the case of non-normalisable wavefunctions the expression simply diverges. It is also worth noting the the dependence on the Wilson line $a$ in (\ref{normland}) drops out since the formal integral over infinity is not sensitive to coordinate shifts. Indeed this is why we choose to include the explicit constant factor of $C_{wl}\left(a\right)$ in the definition (\ref{genprehol}).

\subsection{Kaluza-Klein wavefunctions}
\label{sec:wavekk}

This case is defined by the vanishing of $\chi_{\mathrm{local}}$ which means that one of the eigenvalues of $\mathbb{B}$ vanishes. We take the vanishing eigenvalue to be $\lambda_3$ and so $\lambda_1=-\lambda_2$ and we take $\lambda_1$ to be negative. The eigenvectors of $\mathbb{B}$ still take the same form (\ref{eivecb}) but it is simpler to give an explicit expression for the eigenvalues
\be
\lambda_1 = -\left(R^2\left|m_1\right|^2+R^2\left|m_2\right|^2 + \frac14\left|M_{12}+\bar{M}_{21}\right|^2 + \frac14\left|M_{11}+\bar{M}_{11}\right|^2 \right)^{\frac12}\;.
\ee

It is informative to construct the wavefunction from the Landau-level case (\ref{vphi}). Let us define $\vphi_{\mathrm{gauss}}$ as (\ref{vphi}) in the limit $\lambda_3 \rightarrow 0$. We must also take the prefactor function $f=1$ since now there is no Landau-level multiplicity and there is just one vector-like pair of states at each level.\footnote{The absence of an arbitrary holomorphic prefactor can be seen from the fact that now we must impose both $\tilde{D}_1^- f=0$ and $\tilde{D}_3^- f=0$ giving two complex constraints which generically set the coefficients to vanish apart from the constant term.} Since now $\tilde{D}_3^{\pm}$ commute this wavefunction is annihilated by $\tilde{D}_1^-$, $\tilde{D}_2^+$ and $\tilde{D}_3^{\pm}$. This is the wavefunction for vanishing KK momentum, while the most general solution should satisfy (\ref{kkphieq}). We can write this as $\vphi=\vphi_{\mathrm{gauss}} \vphi_{\mathrm{KK}}$ and require
\be
\tilde{D}_1^-\vphi_{\mathrm{KK}}=\tilde{D}_2^+\vphi_{\mathrm{KK}}=0\;,\;\; \tilde{D}_3^{\pm}\vphi_{\mathrm{KK}} = -i k_K^{\pm}\vphi_{\mathrm{KK}}\;.
\ee 
Solving these equation then gives the full wavefunction
\be
\varphi = e^{-p_1 |z_1|^2 - p_2 |z_2|^2 + p_3 \bar{z}_1z_2 + p_4 \bar{z}_2 z_1 + a_1^K z_1 + a_2^K z_2 + a_3^K \bar{z}_1 +a_4^K \bar{z}_2} \;,
\ee
with 
\bea
a^K_1 &=& -\frac{i\bar{k}_K \hx_{1,2}}{\hx_{3,1} \hx_{1,2} - \hx_{3,2}\hx_{1,1}} \;, \;\;\;\;\;  
a^K_2 = \frac{i\bar{k}_K \hx_{1,1}}{\hx_{3,1} \hx_{1,2} - \hx_{3,2}\hx_{1,1}} \;, \nn \\
\bar{a}^K_3 &=& \frac{i\bar{k}_K \hx_{2,2}}{\hx_{3,1} \hx_{2,2} - \hx_{3,2}\hx_{2,1}} \;, \;\; 
\bar{a}^K_4 = -\frac{i\bar{k}_K \hx_{2,1}}{\hx_{3,1} \hx_{2,2} - \hx_{3,2}\hx_{2,1}} \;\;. \label{akkk}
\eea
Note that the $\vphi_{KK}$ part of the wavefunction is not in general a pure phase. It reduces to a phase if the flux vanishes which forms a subset of the cases where $\chi_{\mathrm{local}}$ vanishes. Nonetheless the amplitude of the wavefunction always becomes flat along an isometry direction as shown in section \ref{sec:locspe}.

Because of this isometry the associated normalisation integral diverges. It is still possible to integrate the wavefunctions in the direction orthogonal to the matter curve and what remains in the integral is just the volume of the matter curve. Performing this gives the normalisation expression
\be
N^2 = \frac{\pi R_{\parallel}^2 \;\mathrm{Vol\;}{\cal C}}{n_1}e^{\frac{|n_2|^2}{4n_1}}\;, \label{normkknofl}
\ee 
where we have defined
\bea
n_1 &\equiv& \left(p_1+\bar{p}_1\right) \left|\hx_{3,2}\right|^2 + \left(p_2+\bar{p}_2\right)\left|\hx_{3,1}\right|^2 
+ \left(p_3+\bar{p}_4\right)\hx_{3,2}\hx^*_{3,1} + \left(p_4+\bar{p}_3\right)\hx_{3,1}\hx^*_{3,2} \;, \nn \\
n_2 &\equiv& -\hx^*_{3,2}\left(a_1^K+\bar{a}_3^K\right) + \hx^*_{3,1}\left(a_2^K+\bar{a}_4^K\right)\;.
\eea
Here $\mathrm{Vol\;}{\cal C}$ is the volume of the matter curve which for a near homogeneous $S_{GUT}$ is approximately given by $\mathrm{Vol\;}{\cal C}\sim \pi R_{\parallel}^2$. This expression where the wavefunction is completely delocalised along the matter curve should form the upper bound on the normalisation prefactor. If the localisation along the curve is sufficiently strong such that the expression (\ref{normland}) gives a smaller value for the normalisation $N$ then it can be trusted, otherwise, if it gives a larger value then the form (\ref{normkknofl}) is more accurate.

It is worth displaying the wavefunction for the special case where the flux vanishes in which case we have
\be
\varphi = e^{-\frac{2R}{\left|m\right|} \left|m_1z_1+m_2z_2\right|^2} \;, \label{wvnoflux}
\ee
where we defined $\left|m\right|^2=\left|m_1\right|^2+\left|m_2\right|^2$. This form is important to keep in mind because we see that any Higgs profile localises the wavefunction onto the matter curve with a strength $\sim R\left|m\right|$. However, as discussed in section \ref{sec:effthe}, the flux need not localise but can also repel the wavefunction away from the matter curve. Although it is not necessary to have the wavefunctions localise to the local patch it is necessary, for the theory to satisfy its purpose, that the operators of the theory, or equivalently the triple overlaps, do localise. This means that we should require that the Higgs localisation dominates over the flux delocalisation including the effects of local Wilson lines. This is essentially automatic within the regime $R \gg 1$ we are working in.\footnote{It is worth noting a small puzzle at this point: because of the finite width of the wavefunctions with respect to the Higgs curve localisation, set by the parameter $R$ in (\ref{wvnoflux}), the scaling of their overlaps differs from the prescribed modular weights found in \cite{Ibanez:1998rf}. For example it is simple to check (see \cite{Camara:2011nj} for a more general expression including massive modes) that the physical Yukawa couplings scale as $R_{\parallel}^{-2} R^{-1/2}$ which differs by the factor of $R^{-1/2}$ from the usual scaling. Due to the difficulty in matching the local scales with global ones and the relation of $R$ to the string coupling, it is unclear to the author whether the different scaling is a physical effect related to finite string coupling or due to a different parameterisation, and whether it persists for $R_{\perp} \gg R_{\parallel}$ where, as discussed in section \ref{sec:eqmot}, the local theory is unreliable. It is also worth noting that in \cite{Kawano:2011aa} it was argued that the kinetic matrix could be evaluated as a residue integral over the matter curves rather than the full $S_{GUT}$ which may be related to the above issue, though the scaling of the integral given in \cite{Kawano:2011aa} with $R$ seems to be the same as would be for an integral over $S_{GUT}$.}

\subsection{GUT singlet wavefunctions}
\label{sec:gutsing}

The GUT singlets are by their nature more difficult to control and quantify within a local theory defined on a patch on $S_{GUT}$. These modes arise from `bulk brane' intersections and so are not restricted to $S_{GUT}$. However the projection of the intersection curve of the bulk branes, on which the singlets are localised, to $S_{GUT}$ is given by the curve on $S_{GUT}$ as defined by the charges of the singlets and the associated Higgs profile $\left<\varphi_H\right>$. Since the charged modes are localised on $S_{GUT}$ in evaluating overlap integrals between charged and singlet modes we expect that replacing the bulk curve with its projection to $S_{GUT}$ should be a decent approximation. Therefore for the purpose of overlap integrals we can solve for the singlet wavefunctions using the general prescription outlines in the previous sections using the appropriate charges under the $S\left(U(1)^5\right)$ group. 

The more serious ambiguity appears in the normalisation integral of the singlets. This integral should not be restricted to $S_{GUT}$ but should go over the full bulk curve. For some calculations involving the GUT singlets the normalisation factor does not play a role. An example of this is in the relation between proton decay and exotic masses as studied in section \ref{sec:exdimfiv}. However more generally the normalisation is important and forms an ambiguity which it seems is difficult to resolve within a local context. 

Practically we expect the normalisation of the singlets to be of the KK form (\ref{normkknofl}) but with
\be
\mathrm{Vol\;}{\cal C} \sim \pi \left(R^{\mathrm{global}}_{\perp}\right)^2
\ee
where $R^{\mathrm{global}}_{\perp}$ is associated to the global version of the local $R_{\perp}$. Typically this is expected to be of order $10$ for local models \cite{Conlon:2009xf} and so we will use $R^{\mathrm{global}}_{\perp} = 10$ for numerical evaluation. However it is important to keep in mind that there is some uncertainty in this prescription for example, if the bulk brane is also local and does not stretch over the full CY volume, or if the flux induces chirality on the singlet curve such that the wavefunction becomes localised within that curve. Therefore results which depend on the normalisation may not be reliable to accuracy of ${\cal O}(10)$ or so.

\subsection{Wavefunction overlaps}
\label{sec:waveover}

Having calculated the form of the wavefunctions we can calculate their triple overlaps which in turn give the associated coefficients of the operators in the four-dimensional theory. The relevant operator in the 8-dimensional theory is given by 
\be
W_Y = \int_S \mathrm{Tr\;} \left[ {\bf A} \wedge {\bf A} \wedge {\bf \Phi} \right]\;. \label{supyuk}
\ee
Thus, having calculated the wavefunctions for the massless and massive charged fields, cubic couplings in the 4-dimensional effective theory are given by integrating the triple overlap of the associated wavefunctions over the internal 4-cycle $S$. Since the wavefunctions are localised within the local patch so is their overlap, which means that effectively we can perform the integral over $S$ as an integral over ${\mathbb C}^2$.

Physical cubic couplings are determined from overlaps of normalised wavefunctions, the normalisation condition assuring 4-dimensional canonically normalised kinetic terms. Note that the canonical normalisation condition is crucial to make sense of the superpotential expression since by itself (\ref{supyuk}) is invariant under the full complexified gauge group \cite{Cecotti:2009zf} which means that it can not hold the full physical information. It is only the combined system of (\ref{supyuk}) and the kinetic terms that is physical.

Wavefunctions for the charged fields can be decomposed as
\be
{\bf A}_{\bar{1}} = \phi_{4D}^I \otimes \psi_{\bar{1}}^I \;,\;\; {\bf A}_{\bar{2}} = \phi_{4D}^I \otimes \psi_{\bar{2}}^I  \;,\;\; {\bf \Phi}_{12} = \phi_{4D}^I \otimes \chi^I \;.
\ee
Here $\phi_{4D}$ are four-dimensional (super-)fields which do not depend on the coordinates on $S_{GUT}$ while the internal profiles are given by $\psi_i$ and $\chi$. The index $I$ runs over all the representations present in the decomposition of the adjoint representation of the full enhanced gauge group $G$ under the remaining gauge group after turning on the Higgs vev and fluxes. The case of $G=E_8$ breaking to the MSSM gauge group is described in section \ref{sec:modbui}. Since we are considering maximal breaking to the Abelian Cartan subgroup commutant with the GUT/MSSM group, the generator structure of the internal wavefunctions is such that the trace in (\ref{supyuk}) simply leads to a selection rule stating that the charges of the three fields under the $U(1)$s should sum to zero. After accounting for this selection rule the relevant four-dimensional cubic coupling is given by
\bea
Y^{\left(I,J,K\right)(i,j,k)}_{p,(n,m,l)} &=& \frac16 G_{IJK} \int_S \left[ \psi^{I,i}_{\bar{1}\, p,(n,m,l)} \psi^{J,j}_{\bar{2}} \chi^{K,k} + \psi^{I,i}_{\bar{2}\, p,(n,m,l)} \psi^{K,k}_{\bar{1}} \chi^{J,j} \right. \nn \\
& & \quad - \chi^{I,i}_{p,(n,m,l)} \psi^{K,k}_{\bar{1}} \psi^{J,j}_{\bar{2}} - \psi^{I,i}_{\bar{1}\, p,(n,m,l)} \psi^{K,k}_{\bar{2}} \chi^{J,j}  \nn \\
& & \quad\left. - \psi^{I,i}_{\bar{2}\, p,(n,m,l)} \chi^{K,k} \psi^{J,j}_{\bar{1}} + \chi^{I,i}_{p,(n,m,l)} \psi^{K,k}_{\bar{2}} \psi^{J,j}_{\bar{1}} \right] \;.
\label{yukexpterms}
\eea
We have split off from the indices $\left\{I,J,K\right\}$ the generation indices $\left\{i,j,k\right\}$ so that generation independent quantities are manifestly so. Hence a four-dimensional field is specified by fixing $I$ and $i$. The coupling is between 3 such canonically normalised fields and we have allowed one of the states in the cubic coupling to be a massive state which is therefore labeled, as in sections \ref{sec:wave} and \ref{sec:wavekk} by the indices $p$, which denotes which of the 3 towers it is in, and $n$, $m$, $l$ which label the excited state. Note that if it is a KK state the index $l$ is replaced with the KK number $k_K$. The indices $i$, $j$ and $k$, denote the possible Landau level degeneracies of the states, i.e. the generation number in the case of multiple generations coming from flux. Finally the factor $G_{IJK}$ accounts for the $U(1)$ selection rules so that for appropriately normalised generators it gives 1 if the coupling is gauge invariant and vanishes if it is not. The triple coupling can be written as
\be
Y^{\left(I,J,K\right)(i,j,k)}_{p,(n,m,l)} = \frac{\left[\hx^I_{p,1} \hx_{1,3}^{[K}\hx_{1,2}^{J]}+\hx^I_{p,2} \hx_{1,1}^{[K}\hx_{1,3}^{J]}+\hx^I_{p,3} \hx_{1,2}^{[K}\hx_{1,1}^{J]} \right] G_{IJK}}{6N_{I,i}N_{J,j}N_{K,k}} \int_S \vphi^{I,i}_{(n,m,l)} \vphi^{J,j} \vphi^{K,k} \;, \label{ygeneral}
\ee
where the square brackets on the indices denote anti-symmetrisation. The excited wavefunction is defined as
\be
\vphi^{I,i}_{(n,m,l)}=\frac{\left(i\tilde{D}_1^+\right)^{n}\left(i\tilde D_2^-\right)^m\left(i\tilde D_3^-\right)^l }{\sqrt{m!n!l!}\left(-\lambda^I_1\right)^{n/2}\left(\lambda^I_2\right)^{m/2}\left(\lambda^I_3\right)^{l/2}}\;\vphi^{I,i}\;.\label{massiverepphi}
\ee
What remains is to give the explicit expression for the overlap integral. Although the general expression is readily calculable we give two cases, which restrict the possible Landau-level excitations participating in the coupling, that can be written in a relatively compact form. The first case is where all the states are ${\cal N}=1$ Landau-levels and where one of the states may be a massive one with arbitrary excitation along the $\tilde D_2^-$ and $\tilde D_3^-$ operators. If we parameterise the generations by the leading power of the $z_i$, rather than through the local Wilson lines (see section \ref{sec:wavell} for a discussion), there is a selection rule operating such that a triple coupling $Y^{(i,j,k)}_{p,(n,m,l)}$ is generally only non-vanishing if 
\be
i+j+k+n\leq m+l \;, \label{zu1selec}
\ee
otherwise the integral over the wavefunctions vanishes identically. It is possible to see this by assigning $U(1)$ charges to the wavefunctions where each power of $z_i$ gives $+1$ while $\bar{z}_i$ gives $-1$. Unless the total overlap integral has no net such charge it will vanish by the integration over the phases of the $z_i$. Note that generally, as discussed in section \ref{sec:wave}, the charge is assigned according to the leading power of $z_i$ or $\bar{z}_i$, but each wavefunction can also have higher order powers, hence the inequality sign in (\ref{zu1selec}). Non-generally, if the holomorphic functions labeling the generations are polynomials of definite degree, as is the case when the matter curve is topologically $\mathbb{P}^1$, the inequality must be saturated.

The selection rule implies that the leading coupling will be to states with no powers of the $\tilde D_1^+$ creation operator since this would add net powers of $\left|z_i\right|^2$ to the integrand as it has positive $U(1)$ charge. For such couplings the triple overlap is given by
\bea
& &\int_S \vphi^{I,i}_{(0,n,m)} \vphi^{J,j} \vphi^{K,k} = \frac{
i^{n+m}C^I_{wl}C^J_{wl}C^K_{wl}}{\sqrt{m!n!}\left(\lambda^I_2\right)^{n/2}\left(\lambda^I_3\right)^{m/2}}\sum_{k_1=0}^i \sum_{k_2=0}^j \sum_{k_3=0}^k \sum_{k_4=0}^n \sum_{k_5=0}^m 
  \begin{pmatrix} i \\ k_1\end{pmatrix} \begin{pmatrix} j \\ k_2\end{pmatrix} \begin{pmatrix} k \\ k_3\end{pmatrix}
  \begin{pmatrix} n \\ k_4\end{pmatrix} \begin{pmatrix} m \\ k_5\end{pmatrix} \nn \\
& &\left(-\hx^I_{1,2}\right)^{i-k_1} \left(-\hx^J_{1,2}\right)^{j-k_2} \left(-\hx^K_{1,2}\right)^{k-k_3} \left(\hx^I_{1,1}\right)^{k_1} \left(\hx^J_{1,1}\right)^{k_2} \left(\hx^K_{1,1}\right)^{k_3} 
\left(k_1^2\right)^{n-k_4}\left(k_1^3\right)^{m-k_5} \left(k_2^2\right)^{k_4} \left(k_2^3\right)^{k_5} \nn \\
& &I\left(i-k_1+j-k_2+k-k_3,n-k_4+m-k_5,k_1+k_2+k_3,k_4+k_5;\right. \nn \\
& &\left. \;\;\;\;p_1^I+p_1^J+p_1^K,p_2^I+p_2^J+p_2^K,p_3^I+p_3^J+p_3^K,p_4^I+p_4^J+p_4^K;c_1^I+c_1^J+c_1^K,c_2^I+c_2^J+c_2^K,0,0\right) \;,  \label{phi3van}
\eea
where $\begin{pmatrix} i \\ k_1\end{pmatrix}$ denote binomial coefficients and we have introduced the expressions
\bea
k_1^i &=& \hx_{i,1}^I \left(-p_1^I - \frac12 \bar{M}_{11}^I \right) + \hx_{i,2}^I \left(p_3^I - \frac12 \bar{M}_{12}^I \right) - \hx^I_{i,3} R \bar{m}_1^I \;, \nn \\
k_2^i &=& \hx_{i,1}^I \left(p_4^I - \frac12 \bar{M}_{21}^I \right) + \hx_{i,2}^I \left(-p_2^I - \frac12 \bar{M}_{22}^I \right) - \hx^I_{i,3} R \bar{m}_2^I \;, \label{k12ll}
\eea
which are for $i=2,3$. Note that in (\ref{phi3van}) we have taken the inequality (\ref{zu1selec}) to be saturated. If it is not saturated there is an arbitrary multiplicative factor which corresponds to the appropriate, non-leading, term in the expansion of the wavefunctions as discussed in section \ref{sec:wave}. Generically this factor would be of order 1.

In the case where the possibly heavy mode is of ${\cal N}=2$ Kaluza-Klein type the overlap integral takes the form
\bea
& &\int_S \vphi^{I}_{(n,0,k_K)} \vphi^{J,j} \vphi^{K,k} =  \frac{i^nC^J_{wl}C^K_{wl}}{\sqrt{n!}\left(-\lambda^I_1\right)^{n/2}}\sum_{k_2=0}^j \sum_{k_3=0}^k \sum_{k_4=0}^n \begin{pmatrix} j \\ k_2\end{pmatrix} \begin{pmatrix} k \\ k_3\end{pmatrix} \begin{pmatrix} n \\ k_4\end{pmatrix} \\ & &
\left(-\hx^J_{1,2}\right)^{j-k_2} \left(-\hx^K_{1,2}\right)^{k-k_3} \left(\hx^J_{1,1}\right)^{k_2} \left(\hx^K_{1,1}\right)^{k_3}
\left(k_1^1\right)^{n-k_4} \left(k_2^1\right)^{k_4} \nn \\
& & I\left(j-k_2+k-k_3+n-k_4,0,k_2+k_3+k_4,0;p_1^I+p_1^J+p_1^K,p_2^I+p_2^J+p_2^K, \right. \nn \\
& & \left. \;\;\;\;p_3^I+p_3^J+p_3^K,p_4^I+p_4^J+p_4^K; \left(a_1^K\right)^I+c_1^J+c_1^K,\left(a_2^K\right)^I+c_2^J+c_2^K,\left(a_3^K\right)^I,\left(a_4^K\right)^I\right) \;, \nn
\eea
where the expressions for the $a_i^K$ are given in (\ref{akkk}). We have presented the expression allowing only a single Landau-level creation operator to act since the more general expression is more cumbersome as the two Landau-level creation operators do not commute. The analogous expression for the other Landau-level operator reads
\bea
& &\int_S \vphi^{I}_{(0,n,k_K)} \vphi^{J,j} \vphi^{K,k} = \frac{i^nC^J_{wl}C^K_{wl}}{\sqrt{n!}\left(\lambda^I_2\right)^{n/2}}\sum_{k_2=0}^j \sum_{k_3=0}^k \sum_{k_4=0}^n \begin{pmatrix} j \\ k_2\end{pmatrix} \begin{pmatrix} k \\ k_3\end{pmatrix} \begin{pmatrix} n \\ k_4\end{pmatrix} \nn \\ & &
\left(-\hx^J_{1,2}\right)^{j-k_2} \left(-\hx^K_{1,2}\right)^{k-k_3} \left(\hx^J_{1,1}\right)^{k_2} \left(\hx^K_{1,1}\right)^{k_3}
\left(k_1^2\right)^{n-k_4} \left(k_2^2\right)^{k_4} \nn \\
& & I\left(j-k_2+k-k_3,n-k_4,k_2+k_3,k_4;p_1^I+p_1^J+p_1^K, p_2^I+p_2^J+p_2^K,p_3^I+p_3^J+p_3^K,p_4^I+p_4^J+p_4^K; \right.\nn \\
& & \left. \left(a_1^K\right)^I+c_1^J+c_1^K,\left(a_2^K\right)^I+c_2^J+c_2^K,\left(a_3^K\right)^I,\left(a_4^K\right)^I\right) \;. \nn
\eea
It is worth noting that in the case with non-vanishing KK momentum $k_K$ the selection rule (\ref{zu1selec}) no longer holds for these couplings.

Another type of cubic coupling that arises is a coupling to the conjugate representations ${\bf \bar{R}}$. In the case of Landau-Level modes these are, by convention, always massive (\ref{rrbmass}). Nonetheless these couplings are important when integrating out massive modes. We will consider cubic couplings where the first mode, which as above may be an excited state, is in the ${\bf \bar{R}}$ representation. As studied in section \ref{sec:n1spec}, this amounts to taking the complex conjugate wavefunction (\ref{rrbcc}) so that we are interested in, for the Landau-Levels, the integral
\be
\int_S \left(\vphi^{I,i}_{(n,m,l)}\right)^* \vphi^{J,j} \vphi^{K,k} \;. \label{rbarcubicgen}
\ee
In this case the presence of an inequality analogous to (\ref{zu1selec}) depends on the form of the holomorphic functions $f_l$. As discussed in section \ref{sec:wave}, depending on the global topology of the matter curve this can take the form of an exact power of the $z_i$, or it may be some series starting with a fixed power but including higher order terms. In the latter case, since the first term in (\ref{rbarcubicgen}) includes potentially all powers of $z$ and $\bar{z}$, there is no selection rule at all. In the former case we have the selection rule
\be
i+n=m+l+j+k\;. \label{lbin}
\ee
The coupling (\ref{rbarcubicgen}) evaluates to
\bea
& &\int_S \left(\vphi^{I,i}_{(n,m,l)}\right)^* \vphi^{J,j} \vphi^{K,k} = \frac{
\alpha \left(-i\right)^{n+m+l}\bar{C}^I_{wl}C^J_{wl}C^K_{wl}}{\sqrt{m!n!l!}\left(-\lambda^I_1\right)^{n/2}\left(\lambda^I_2\right)^{m/2}\left(\lambda^I_3\right)^{l/2}} \nn \\
& &\sum_{k_1=0}^i \sum_{k_2=0}^j \sum_{k_3=0}^k \sum_{k_4=0}^n \sum_{k_5=0}^m \sum_{k_6=0}^l 
  \begin{pmatrix} i \\ k_1\end{pmatrix} \begin{pmatrix} j \\ k_2\end{pmatrix} \begin{pmatrix} k \\ k_3\end{pmatrix}
  \begin{pmatrix} n \\ k_4\end{pmatrix} \begin{pmatrix} m \\ k_5\end{pmatrix} \begin{pmatrix} l \\ k_6\end{pmatrix} \nn \\
& &\left(-\left(\hx^I_{1,2}\right)^*\right)^{i-k_1} \left(-\hx^J_{1,2}\right)^{j-k_2} \left(-\hx^K_{1,2}\right)^{k-k_3} \left(\left(\hx^I_{1,1}\right)^*\right)^{k_1} \left(\hx^J_{1,1}\right)^{k_2} \left(\hx^K_{1,1}\right)^{k_3} \nn \\
& &\left(\bar{k}_1^1\right)^{n-k_4}\left(\bar{k}_1^2\right)^{m-k_5}\left(\bar{k}_1^3\right)^{l-k_6} \left(\bar{k}_2^1\right)^{k_4}\left(\bar{k}_2^2\right)^{k_5} \left(\bar{k}_2^3\right)^{k_6} \nn \\
& &I\left(j-k_2+k-k_3+m-k_5+l-k_6,i-k_1+n-k_4,k_2+k_3+k_5+k_6,k_1+k_4;\right. \nn \\
& & \;\;\;\;\bar{p}_1^I+p_1^J+p_1^K,\bar{p}_2^I+p_2^J+p_2^K,\bar{p}_3^I+p_3^J+p_3^K,\bar{p}_4^I+p_4^J+p_4^K; \nn \\
& & \left. \;\;\;\; c_1^J+c_1^K,c_2^J+c_2^K,\bar{c}_1^I,\bar{c}_2^I \right) \;,  \label{phi3rbll}
\eea
Here we have introduced
\bea
k_1^1 &=& \left(\hx_{1,1}^*\right)^I \left(-p_1^I + \frac12 M_{11}^I \right) + \left(\hx_{1,2}^*\right)^I \left(p_4^I + \frac12 M_{12}^I \right) + \left(\hx^*_{1,3}\right)^I R m_1^I \;, \nn \\
k_2^1 &=& \left(\hx_{1,1}^*\right)^I \left(p_3^I + \frac12 M_{21}^I \right) + \left(\hx_{1,2}^*\right)^I \left(-p_2^I + \frac12 M_{22}^I \right) + \left(\hx^*_{1,3}\right)^I R m_2^I \;, \label{k12ll2}
\eea
The extra coefficient $\alpha$ accounts for the fact that there are two possibilities of indices values that lead to non-vanishing results. The first is when the equality (\ref{lbin}) is satisfied, in which case $\alpha=1$, and the second is where it is not satisfied and there are higher order terms in the holomorphic functions, a possibility that depends on the global topology of the matter curve, in which case $\alpha$ is the factor multiplying the appropriate higher order term.\footnote{There is a similar factor multiplying the expression (\ref{phi3van}) which we dropped because it plays a less important role there.}

There are also analogous expressions for coupling to KK modes which read
\bea
& &\int_S \left(\vphi^{I}_{(n,0,k_K)}\right)^* \vphi^{J,j} \vphi^{K,k} =  \frac{\left(-i\right)^nC^J_{wl}C^K_{wl}}{\sqrt{n!}\left(-\lambda^I_1\right)^{n/2}}\sum_{k_2=0}^j \sum_{k_3=0}^k \sum_{k_4=0}^n \begin{pmatrix} j \\ k_2\end{pmatrix} \begin{pmatrix} k \\ k_3\end{pmatrix} \begin{pmatrix} n \\ k_4\end{pmatrix} \\ & &
\left(-\hx^J_{1,2}\right)^{j-k_2} \left(-\hx^K_{1,2}\right)^{k-k_3} \left(\hx^J_{1,1}\right)^{k_2} \left(\hx^K_{1,1}\right)^{k_3}
\left(\bar{k}_1^1\right)^{n-k_4} \left(\bar{k}_2^1\right)^{k_4} \nn \\
& & I\left(j-k_2+k-k_3,n-k_4,k_2+k_3,k_4;\bar{p}_1^I+p_1^J+p_1^K,\bar{p}_2^I+p_2^J+p_2^K,\bar{p}_3^I+p_3^J+p_3^K,\bar{p}_4^I+p_4^J+p_4^K; \right. \nn \\
& & \left. \;\;\;\; \left(\bar{a}_1^K\right)^I+c_1^J+c_1^K,\left(\bar{a}_2^K\right)^I+c_2^J+c_2^K,\left(\bar{a}_3^K\right)^I,\left(\bar{a}_4^K\right)^I\right) \;, \nn
\eea
and similarly for the other tower
\bea
& &\int_S \left(\vphi^{I}_{(0,n,k_K)}\right)^* \vphi^{J,j} \vphi^{K,k} = \frac{\left(-i\right)^nC^J_{wl}C^K_{wl}}{\sqrt{n!}\left(\lambda^I_2\right)^{n/2}}\sum_{k_2=0}^j \sum_{k_3=0}^k \sum_{k_4=0}^n \begin{pmatrix} j \\ k_2\end{pmatrix} \begin{pmatrix} k \\ k_3\end{pmatrix} \begin{pmatrix} n \\ k_4\end{pmatrix}  \\ & &
\left(-\hx^J_{1,2}\right)^{j-k_2} \left(-\hx^K_{1,2}\right)^{k-k_3} \left(\hx^J_{1,1}\right)^{k_2} \left(\hx^K_{1,1}\right)^{k_3}
\left(\bar{k}_1^2\right)^{n-k_4} \left(\bar{k}_2^2\right)^{k_4} \nn \\
& & I\left(j-k_2+k-k_3+n-k_4,0,k_2+k_3+k_4,0;\bar{p}_1^I+p_1^J+p_1^K, \bar{p}_2^I+p_2^J+p_2^K,\bar{p}_3^I+p_3^J+p_3^K, \right. \nn \\
& & \left. \bar{p}_4^I+p_4^J+p_4^K; \left(\bar{a}_1^K\right)^I+c_1^J+c_1^K,\left(\bar{a}_2^K\right)^I+c_2^J+c_2^K,\left(\bar{a}_3^K\right)^I,\left(\bar{a}_4^K\right)^I\right) \;. \nn
\eea

There are also cubic couplings that involve two massive modes and one massless one. These also play a role, for example in generating dimension 5 proton decay operators in the presence of a $U(1)_{PQ}$ symmetry as studied in section \ref{sec:exdimfiv}. For the sake of brevity we will not display the wavefunction overlap expressions for all the combinations of massive modes. They can be calculated in the same way as above. Instead we present an example overlap which is the one used in section \ref{sec:exdimfiv}. This involves a cubic coupling between two conjugate massive KK modes (with no LL excitations) and a massless mode. The appropriate overlap reads
\bea
& &\int_S \left(\vphi^{I}_{(0,0,k^I_K)}\right)^* \left(\vphi^{J}_{(0,0,k^J_K)}\right)^* \vphi^{K,k} =  C^K_{wl}\sum_{k_3=0}^k \begin{pmatrix} k \\ k_3\end{pmatrix}  \left(-\hx^K_{1,2}\right)^{k-k_3} \left(\hx^K_{1,1}\right)^{k_3}
\left(\bar{k}_1^1\right)^{n-k_4} \left(\bar{k}_2^1\right)^{k_4} \nn \\
& & I\left(k-k_3,0,k_3,0; \bar{p}_1^I+\bar{p}_1^J+p_1^K,\bar{p}_2^I+\bar{p}_2^J+p_2^K,\bar{p}_3^I+\bar{p}_3^J+p_3^K,\bar{p}_4^I+\bar{p}_4^J+p_4^K; \right.\nn \\ 
& & \left. \left(\bar{a}_1^K\right)^I+\left(\bar{a}_1^K\right)^J+c_1^K,\left(\bar{a}_2^K\right)^I+\left(\bar{a}_2^K\right)^J+c_2^K,\left(\bar{a}_3^K\right)^I+\left(\bar{a}_3^K\right)^J,
\left(\bar{a}_4^K\right)^I+\left(\bar{a}_4^K\right)^J \right) \;.
\eea
Since there are two massive modes involved in the coupling we should generalise the prefactor in the expression (\ref{ygeneral}) so that if we associate the ${\cal N}=4$ tower index $p$ to the field $I$ and $q$ to $J$ we have the prefactor
\be
\hx^I_{p,1} \hx_{1,3}^{K}\hx_{q,2}^{J} - \hx^I_{p,1} \hx_{q,3}^{J}\hx_{1,2}^{K} + \hx^I_{p,2} \hx_{1,1}^{K}\hx_{q,3}^{J} - \hx^I_{p,2} \hx_{q,1}^{J}\hx_{1,3}^{K} + \hx^I_{p,3} \hx_{1,2}^{K}\hx_{q,1}^{J} - \hx^I_{p,3} \hx_{q,2}^{J}\hx_{1,1}^{K}\;.
\ee

\section{Towards model building on the point of $E_8$}
\label{sec:modbui}

The expressions for wavefunctions and their overlaps presented in the previous sections are quite general and hold for any gauge group. In this section we apply these results to the case of a point of $E_8$ enhancement within an $SU(5)$ gauge group over $S_{GUT}$ which is subsequently broken to $SU(3)\times SU(2)\times U(1)_Y$ using hypercharge flux. As outlined in the introduction such a setup corresponds to a particularly rich theory on a local patch. The gauge theory on $S_{GUT}$ has a gauge group $G=E_8$ which is broken to $E_8 \supset SU(5)_{GUT}\times SU(5)_{\perp} \rightarrow SU(5)_{GUT} \times U(1)^4$ by a spatially varying background Higgs vev. To see the matter content of the theory we can decompose the adjoint representation of $E_8$ under $SU(5)_{GUT} \times SU(5)_{\perp}$ as
\be
\bf{248} \rightarrow \left(\bf{24},\bf{1}\right)\op\left(\bf{1},\bf{24}\right)\op\left(\te,\f\right)\op\left(\fb,\te\right)\op\left(\teb,\fb\right)\op\left(\f,\teb\right) \;,
\ee
we see that we have 5 $\te$-matter curves, 10 $\f$-matter curves and 24 singlets. Upon breaking to the Cartan of $SU(5)_{\perp}$ the charges of the fields can be most concisely written in terms of 5-component vectors $\left(\bf q_i\right)_j=\delta_{ij}$ which denote the charge under the 5 $U(1)$s inside $U(1)^4\simeq S\left( U(1)^5 \right)$. Then the different curves can be labeled by their charges as
\bea
\Sigma_{\te} &:& {\bf q_i}  \;, \nn \\
\Sigma_{\f} &:& -{\bf q_i} - {\bf q_j}, \;\; i\neq j\;, \nn \\
\Sigma_{\bf{1}} &:& \left({\bf q_i} - {\bf q_j}\right) , \;\; i\neq j\;, \label{cur3}
\eea
with $i=1,...,5$.
To calculate the wavefunctions for a given matter curve we contract the associated charge vectors with the background Higgs and flux (\ref{phi}) and (\ref{a}) which gives the effective Higgs and flux felt by the modes on the matter curve as in (\ref{efffluxhiggs}). The resulting wavefunctions overlaps are then calculated as in section \ref{sec:waveover} which give the coefficient of the appropriate cubic coupling in the four-dimensional theory. 

Therefore in specifying the local theory the input parameters are the background Higgs $m_i^a$, flux $M_{ij}^a$ along the $S\left( U(1)^5 \right)\times U(1)_Y$ generators, the geometric quantities $R_{\parallel}$ and $R_{\perp}$, and the local Wilson line values $c_1$ and $c_2$ for each wavefunction. Once these are specified the local form of the wavefunctions is determined and the resulting overlaps can be calculated. As well as determining the overlaps, as discussed in section \ref{sec:locspe}, in the absence of local Wilson lines the local values of the fluxes are connected to the chirality of the localised modes.

The variety of possible local flux and Higgs values, and the intricate relations between these quantities and the spectrum and operators of the effective four-dimensional theory leads to a rich and complicated phenomenology covering a wide range of phenomenological issues. A comprehensive study of the various possibilities and their phenomenological implications is beyond the scope of this paper. Instead, at this stage, the aim is to introduce the important concepts and provide the necessary tools to study these theories. To this end in this section we will study an example model which is not chosen for particularly attractive phenomenological properties but rather simply to allow for explicit, rather than abstract, discussions of the type of analyses that can be performed for these models.

Some tools for analysing such models are provided at:
\begin{center}
  \url{http://www.cpht.polytechnique.fr/cpht/palti/pointE8.html}
\end{center}
The web page takes the background values for the flux and Higgs vevs as input values. 
It applies (\ref{locchi}) to the different effective values for all the representations calculating the resulting chiral spectrum in the local patch ignoring the possible local Wilson lines, i.e. it provides the chirality induced in the case where the Wilson lines vanish so that we can be sure that the local patch captures this correctly. It also calculates the effective Higgs and flux felt by each representation according to its charges and there is a link to a Mathematica worksheet which calculates the resulting wavefunctions and overlaps for 3 such matter curves.

A condition which we impose on the Higgs field vev is that locally the gauge rank does not enhance beyond 1 rank except at the point of $E_8$ at the origin. This is not a necessary condition for general models, for example it is possible to turn off matter curves all together, or to enhance by rank 2 over a matter curve. However for simplicity we would like to require the above property, which also ensures that the triple overlap of any wavefunctions, in the presence of just a Higgs background, is localised at the origin. There are of course many values of Higgs fields that satisfy this property but a simple way to ensure it is to take the following ansatz
\be
m_1^a = e^{2 \pi i a / 5} \;,\;\; m_2^a \in \mathrm{Rationals} \neq 0\;.
\ee
Since the fifth roots of unity form a cyclotomic extension of rank 4 over the rationals it is not possible to set equal any two curves for rational non-vanishing values of the $m_2^a$ (other than using the single linear relation which corresponds to the tracelessness constraint). Of course in practice there may be approximate cancellations which could lead to near vanishing Higgs profiles along some directions we should be aware of.

The chosen example model is defined by the local values for the Higgs and Flux backgrounds given in table \ref{examlocval}.
\begin{table}
\centering
\begin{tabular}{|c|c|c|c|}
\hline
Higgs & Value & Flux & Value \\
\hline
$m_1^1$ & $1$ & $m_2^4$ & $-\frac45$ \\
\hline
$m_1^2$ & $e^{2\pi i/5}$ & $m_2^5$ & $-1$ \\
\hline
$m_1^3$ & $e^{4\pi i/5}$ & $M_{11}^1$ & $-\frac{11}{5}$ \\
\hline
$m_1^4$ & $e^{6\pi i/5}$ & $M_{11}^2$ & $\frac{14}{5}$ \\
\hline
$m_1^5$ & $e^{8\pi i/5}$ & $M_{11}^3$ & $-\frac{11}{5}$ \\
\hline
$m_2^1$ & $-\frac35$ & $M_{11}^5$ & $\frac85$ \\
\hline
$m_2^2$ & $\frac75$ & $M_{11}^Y$ & $-\frac95$ \\
\hline
$m_2^3$ & $1$ & & \\
\hline
\end{tabular}
\caption{Table showing the non-vanishing Higgs and flux values for the example model. The super-indices denote the $S\left(U(1)^5\right)\times U\left(1\right)_Y$ generators. Note that the D-terms are implicitly solved so that $M_{22}^a=-M_{11}^a$.}
\label{examlocval}
\end{table}
With the input values specified we go on to discuss aspects of this model as explicit cases of more general properties of these type of local models.

Before proceeding it is worth fixing the local geometric parameters $R_{\perp}$ and $R_{\parallel}$ within the local theory. Given that the Higgs vevs taken are ${\cal O}\left(1\right)$, we choose to study the geometric parameter space by fixing $R_{\perp}=3/2$ and 
$R_{\parallel}=3$ which is consistent with the constraints outlined in section \ref{sec:effthe}. Note that, as emphasised in section \ref{sec:effthe} and in \cite{Camara:2011nj}, the local value for $R_{\perp}$ would have to differ by a factor of ${\cal O}\left(10\right)$ from the global value which sets the Planck scale; this being a universal issue with local gauge theory models.

\subsection{The massless spectrum}
\label{sec:exmassspec}

The first property of the local models to study is the massless spectrum. Since the model is defined on a local patch within $S_{GUT}$ it can not provide full information regarding this but rather is connected to it through the discussion presented in section \ref{sec:locspe}. We choose out local Wilson lines so that the local modes include those given in table \ref{examlocspec}. By definition these modes have vanishing local Wilson lines implying their wavefunctions peak at the origin and their associated local chirality is determined by the background Higgs and fluxes as presented in table \ref{examlocval} applied to the formula (\ref{locchi}). The local chirality for these modes is shown in table \ref{examlocspec}.

\begin{table}
\centering
\begin{tabular}{|c|c|}
\hline
Representation & $\chi_{\mathrm{local}}$  \\
\hline
$\left(\left(3,2\right)_{1/6}\oplus\left(\bar{3},1\right)_{-2/3}\oplus\left(1,1\right)_{1} \right) \otimes {\bf q_1}$ & +1\\
\hline
$\left(\left(3,2\right)_{1/6}\oplus\left(\bar{3},1\right)_{-2/3}\oplus\left(1,1\right)_{1} \right) \otimes {\bf q_2}$ & +1\\
\hline
$\left(\left(3,2\right)_{1/6}\oplus\left(\bar{3},1\right)_{-2/3}\oplus\left(1,1\right)_{1} \right) \otimes {\bf q_3}$ & 0\\
\hline
$\left(\left(\bar{3},1\right)_{1/3}\oplus\left(1,2\right)_{-1/2}\right) \otimes \left(-{\bf q_1}-{\bf q_3}\right)$ & -1\\
\hline
$\left(\left(\bar{3},1\right)_{1/3}\oplus\left(1,2\right)_{-1/2}\right) \otimes \left(-{\bf q_4}-{\bf q_5}\right)$ & -1\\
\hline
$\left(3,1\right)_{-1/3} \otimes \left(-{\bf q_1}-{\bf q_2}\right)$ & 0 \\
\hline
$\left(1,2\right)_{+1/2} \otimes \left(-{\bf q_1}-{\bf q_2}\right)$ & +1 \\
\hline
$\left(3,1\right)_{-1/3} \otimes \left(-{\bf q_2}-{\bf q_3}\right)$ & 0 \\
\hline
$\left(1,2\right)_{+1/2} \otimes \left(-{\bf q_2}-{\bf q_3}\right)$ & -1 \\
\hline
\end{tabular}
\caption{Table showing the chirality for local modes.}
\label{examlocspec}
\end{table}
The local spectrum is compatible with exactly the matter content of the MSSM, with 2 generations living on one of the $\fb$ matter-curve and similarly 2 generations on one of the $\te$ curves. All the other modes may have local Wilson lines and so their chirality is unrelated to the local values of the Higgs and fluxes. However, their local values for $\chi_{\mathrm{local}}$ are non-vanishing which means that although there need not be necessarily any net chirality over the matter curve, or any massless modes at all, we also know that many modes can not be of KK type since that requires $\chi_{\mathrm{local}}=0$ over the full matter curve. 

It is important to note that a motivation for considering models where most of the modes are local is that the presence of their wavefunction peak at the interaction point maximises they size of the coefficient of the operators involving them. This is particularly important in the context of recovering sizable Yukawa couplings.

We would like to embed the chirality of the local spectrum into a global model. We can realise this, at least within the semi-local approach, by considering the semi-local spectrum presented in table \ref{examglobspec}.
\begin{table}
\centering
\begin{tabular}{|c|c|c|}
\hline
Field Name & Representation & Multiplicity  \\
\hline
$\left(Q_1\oplus U_1 \oplus E_1 \right)$ & $\left(\left(3,2\right)_{1/6}\oplus\left(\bar{3},1\right)_{-2/3}\oplus\left(1,1\right)_{1} \right) \otimes {\bf q_1}$ & 1\\
\hline
$\left(Q_{2}\oplus U_{2} \oplus E_{2} \right)$ & $\left(\left(3,2\right)_{1/6}\oplus\left(\bar{3},1\right)_{-2/3}\oplus\left(1,1\right)_{1} \right) \otimes {\bf q_2}$ & 2\\
\hline
$\left(D_{1} \oplus L_{1} \right)$ & $\left(\left(\bar{3},1\right)_{1/3}\oplus\left(1,2\right)_{-1/2}\right) \otimes \left({\bf q_4}+{\bf q_5}\right)$ & 1 \\
\hline
$\left(D_{2,3} \oplus L_{2,3} \right)$ & $\left(\left(\bar{3},1\right)_{1/3}\oplus\left(1,2\right)_{-1/2}\right) \otimes \left({\bf q_1}+{\bf q_3}\right)$ & 2 \\
\hline
$H_U$ & $\left(1,2\right)_{+1/2} \otimes \left(-{\bf q_1}-{\bf q_2}\right)$ & 1 \\
\hline
$H_D$ & $\left(1,2\right)_{-1/2} \otimes \left({\bf q_2}+{\bf q_3}\right)$ & 1 \\
\hline
\hline
$E_1$ &$\left(1,2\right)_{-1/2} \otimes \left({\bf q_1}+{\bf q_5}\right)$ & 1 \\
\hline
$E_2$ &$\left(1,2\right)_{+1/2} \otimes \left(-{\bf q_3}-{\bf q_5}\right)$ & 1 \\
\hline
- &$\left(\left(3,2\right)_{1/6}\oplus2\times\left(1,1\right)_{1} \right) \otimes {\bf q_2}$ & 1\\
\hline
- &$\left(\left(\bar{3},2\right)_{-1/6}\oplus2\times\left(1,1\right)_{-1} \right) \otimes \left(-{\bf q_5}\right)$ & 1\\
\hline
- &$\left(\bar{3},1\right)_{1/3} \otimes \left({\bf q_4}+{\bf q_5}\right)$ & 1 \\
\hline
- &$\left(3,1\right)_{-1/3} \otimes \left(-{\bf q_2}-{\bf q_4}\right)$ & 1 \\
\hline
\hline
$X_1$ & $\left(1,1\right)_0 \otimes \left(-{\bf q_1}+{\bf q_3}\right)$ & 1 \\
\hline
$X_2$ & $\left(1,1\right)_0 \otimes \left(-{\bf q_1}+{\bf q_2}\right)$ & 1 \\
\hline
\end{tabular}
\caption{Table showing the massless spectrum of the example model as determined from a semi-local approach.}
\label{examglobspec}
\end{table}
The presence of additional non-MSSM modes is required by a consistent embedding of the spectrum into $S_{GUT}$. The constraints of this embedding are calculated in the appendix where also the explicit embedding of the spectrum of table \ref{examglobspec} into $S_{GUT}$ is presented. We have also added a massless GUT singlet field $X$.\footnote{The net chirality of the GUT singlets is not constrained in the same way as the charged matter since it is sensitive to the pull back of the flux to divisor intersections away from $S_{GUT}$ and therefore require a full global model to determine. Some local constraints of the chirality of the singlets that arises just from the component of the flux which pulls back to curves on $S_{GUT}$ was studied in \cite{Marsano:2009wr,Callaghan:2011jj}.}
The spectrum of the example model is in many ways typical of the type of models studied in semi-local F-theory GUTs, for example in \cite{Marsano:2009wr,Dudas:2010zb}. It has additional $U(1)$ symmetries with selection rules, in particular this model exhibits a $U(1)_{PQ}$ symmetry which forbids dimension 5 proton decay. It has exotics which are associated to the presence of such a symmetry which can only be lifted by giving a vev to the singlet $X$.\footnote{There are other exotics in table \ref{examglobspec} that are unrelated to the presence of a $U(1)_{PQ}$ which can also be lifted by giving appropriate singlets a vev.} An important aspect by which it differs from more advanced F-theory models is the rank of the up-type Yukawa matrix being 2 rather than 1 because to achieve a rank 1 matrix requires local monodromies \cite{Hayashi:2009ge,Cecotti:2010bp}.

Having defined the massless spectrum of the local theory we can calculate the associated operators. Over the upcoming sections we do so for various type of operators that are of phenomenological interest. It is important to emphasise that primarily due to an ambiguity in the normalisation integral of the wavefunctions, as discussed in section \ref{sec:wave}, the numbers quoted are only expected to hold to within factors of ${\cal O}(1)$ (or perhaps ${\cal O}(10)$).

\subsection{Yukawa couplings}
\label{sec:exayuk}

In this section we study various aspects of Yukawa couplings. We begin by considering the Yukawa coupling to $H_U$, for example $Y_{H_UQ_1U_2}$. All of the fields participating in this coupling are localised within the patch and so this coupling serves as interesting case for studying the dependence of the magnitude of couplings on the approximations used in the local effective theory. Using the formula for the integral (\ref{phi3van}) gives the coupling
\be
Y_{H_UQ_1U_2} \sim 3\times 10^{-3}\;, \label{topyuk}
\ee
where here, and henceforth, we quote the absolute value of the coupling. Our conventions for the normalisation of the wavefunctions is that, for a given wavefunction, out of the Landau-Level normalisation (\ref{normland}) and the Kaluza-Klein normalisation (\ref{normkknofl}) we use whichever one gives the smallest value for the normalisation factor $N$. The motivation being that if the flux localises the wavefunctions significantly within matter curve then the LL normalisation will be more accurate than the KK one which is approximate.

It is clear that the magnitude of the coupling (\ref{topyuk}) is far too small to reproduce the top quark Yukawa coupling which must be close to $1$ at the GUT scale. Since the kinetic terms are canonically normalised there is no overall scale that can be used to reconcile the two values. This problem is quite general and is associated to the difference in localisation of the wavefunctions perpendicular and parallel to the curves which means that the wavefunctions normalisation integral is too large compared with the overlap integral. Generally couplings of ${\cal O}\left(1\right)$ are associated with the large flux limit, $M_{ij} \gg 1$, which implies that the wavefunctions become strongly localised withing the curve, or the geometric limit $R_{\parallel} \ll 1$. Neither of these limits are compatible with the validity of the local theory, however it is plausible that for some regions in the valid local parameters space that are closer to this limit a more substantial top Yukawa coupling can be induced.

The down-type Yukawa coupling is calculated to be
\be
Y_{H_DQ_1D_1} = 3\times 10^{-3}\;. \label{botyuk}
\ee
Note that an important aspect of studying the point of $E_8$ is that, as we have shown, both the up-type and down-type Yukawa couplings can be calculated within the same local theory allowing for a determination of their relation and consequently parameters such as tan$\beta$.

\subsubsection*{Non-normalisable Yukawa couplings}

The generation of Yukawas for the lighter generations is less clearly understood. The reason being that the presence of a normalisable Yukawa coupling for one generation, associated with at least a rank 2 enhancement of the gauge group over a point, does not imply non-vanishing Yukawas for the lighter generations \cite{Cecotti:2009zf,Conlon:2009qq,Marchesano:2009rz}. Therefore some additional features are required to generate the other Yukawas and these depend on how the generations are realised. In the case where the generations are Landau-level multiplicities coming from a single matter curve, the most natural possibility is that there are a number of intersection points along the curve so that at each one a Yukawa coupling is induced. This possibility is the most natural from a top-down perspective, however it is less natural when we account for the observed hierarchies of quark and lepton masses in the SM. It is not clear how and why this should naturally arise within such a construction. Nonetheless it is a viable possibility and we refer to \cite{Hayashi:2009bt} for further studies of this scenario.\footnote{Note that such a scenario could be studied within a local patch though the distance between the intersection points would push the viability of the local theory to the edge.}

Another possibility, studied in \cite{Cecotti:2009zf,Marchesano:2009rz,Aparicio:2011jx}, is that non-perturbative effects induce a non-commutative deformation of the local theory which does induce a rank 3 Yukawa coupling associated to a single enhancement point. This itself has difficulties particularly in embedding into a global realisation, and in generating Yukawas that are not too small for the lighter generations given that they have a non-perturbative origin.

A qualitatively different approach is to associate each of the generations to a different matter curve. This way it is possible to control the structure of Yukawa couplings using the additional $U(1)$ symmetries naturally present in breaking $E_8$ to $SU(5)$ \cite{Dudas:2009hu,King:2010mq,Leontaris:2010zd}. In this case some of the Yukawas arise as non-renormalisable operators in the four-dimensional effective theory after integrating out heavy modes. Such a setup is particularly interesting to study within the local context since all the important couplings can be calculated. For the rest of this section we will study this in detail since it is also relevant for generating an approximate rank 1 up-type Yukawa coupling, as opposed to the rank 2 in the example model. 
\footnote{It is important to note that as we have seen with the top Yukawa the degree and form of localisation plays an important role in the magnitude of the coupling which naturally raises the possibility that the Yukawa structure could arise from the localisation properties of the generations. This has also been suggested in a large number of other works, see \cite{Binetruy:1994bn,Donagi:2011dv} for old and recent references. Indeed we have seen that an accurate determination of the top Yukawa within a local theory hints at point-like localisation for some of the wavefunctions involved. Either way the localisation properties of the wavefunctions clearly play an important role in determining the magnitude of the operators and it certainly deserves further study within local theories where the associated parameters can be more accurately determined.}

In the example model of table \ref{examglobspec} a non-renormalisable Yukawa coupling which is gauge invariant is given by
\be
Y^{NR} X_2 H_U Q_1 U_1 \;. \label{ynr}
\ee
This is an interesting case as it amounts to a four-dimensional version of the monodromy structure typically used in F-theory GUTs to generate a rank 1 up-type Yukawa matrix. The coupling is induced by integrating out massive modes from the renormalisable couplings
\be
\hat{Y}_{X_2Q_1\bar{Q}_2}X_2 Q_1 \bar{Q}_2^{LL} + \hat{Y}_{H_UQ_2U_1}H_U Q_2^{LL} U_1 \;, \label{normyukle}
\ee
where the superscript $LL$ denotes massive Landau-Level states as studied in section (\ref{sec:n1spec}). Because of the fact that the mass operator of massive LL also changes the level and so the wavefunction these couplings are non-trivial in the sense that they require an understanding of the holomorphic part of the wavefunctions beyond the leading expressions (\ref{polyexp}). This is because, as seen in the model of table \ref{examglobspec}, there is only one generation on each of the curves carrying $H_U$, $Q_1$ and $U_1$ which means that the leading parts of the holomorphic prefactor of the wavefunctions is constant. However this part can not contribute to the coupling (\ref{ynr}) because in (\ref{normyukle}) either $Q_2^{LL}$ or $\bar{Q}_2^{LL}$ must have a non-trivial dependence on net powers of $z_i$ which would mean that the appropriate cubic wavefunction integral would vanish.

Therefore, in evaluating a non-vanishing contribution to (\ref{ynr}) from integrating out massive LL, there is an ambiguity relating to the coefficient of the subleading holomorphic power in the wavefunction. As discussed in section \ref{sec:wave}, there are two unknown factors in determining this coefficient, the first being if there are subleading powers present at all which depends on the topology of the matter curve. For the example studied in section \ref{sec:locspe} of $\mathbb{P}^1$ there were no such powers as they would lead to non-normalisable wavefunctions, while in the case of a torus such higher order terms are present. Assuming that such terms are present there is then an arbitrary coefficient multiplying them which only gives a sub-leading contribution to the normalisation integral and so is relatively unconstrained though expected to be of order one. We will calculate the contribution to (\ref{ynr}) that arises from a linear holomorphic term in the wavefunction prefactor of $X_2$. This is expected to be present since the wavefunction for $X_2$ is the pullback of a bulk wavefunction to the GUT brane anyway and so is expected to have higher order powers, further since we have little control of the normalisation of the singlet wavefunction and have to estimate it the ambiguity in the coefficient of the linear holomorphic term is absorbed into the normalisation. Therefore, we should keep in mind that the result can only be trusted to within factors of ${\cal O}\left(10\right)$ or so.

From the grouping of LL calculated in (\ref{llgrps}) we see that the appropriate heavy modes to integrate out are the pairings
\be
\left\{Q_2^{LL},\bar{Q}_2^{LL}\right\} = \left\{\Psi^{\bf R}_{2,(0,0,0)},\Psi^{\bf \bar{R}}_{3,(1,0,0)}\right\} \;, \left\{\Psi^{\bf R}_{3,(0,0,0)},\Psi^{\bf \bar{R}}_{2,(1,0,0)}\right\}\;\; \;. \label{yuknnhm}
\ee
These are expected to give the leading contribution to the non-normalisable operator. Labeling the coupling to the first set of heavy modes with superscript 1 and to the second with superscript 2, taking the coefficient of the linear term in the singlet wavefunction to be 1 and using the approximate singlet normalisation discussed in section \ref{sec:gutsing}, the formulas presented in section \ref{sec:waveover} give
\bea
\hat{Y}^1_{X_2Q_1\bar{Q}_2} &\simeq& 7\times 10^{-5}\;,\;\; \hat{Y}^1_{H_UQ_2U_1} \simeq 5\times 10^{-4}\;, \nn \\
\hat{Y}^2_{X_2Q_1\bar{Q}_2} &\simeq& 10^{-4}\;,\;\; \hat{Y}^2_{H_UQ_2U_1} \simeq 10^{-3}\;.
\eea
The contribution from the heavy modes (\ref{yuknnhm}) to the coupling (\ref{ynr}) is therefore calculated to be
\be
Y^{NR} \simeq \frac{1}{M_{K}\sqrt{-\lambda_1}} \left(\hat{Y}^1_{X_2Q_1\bar{Q}_2} \hat{Y}^1_{H_UQ_2U_1} + \hat{Y}^2_{X_2Q_1\bar{Q}_2} \hat{Y}^2_{H_UQ_2U_1} \right) \sim \frac{3\times 10^{-8}}{M_{K}} \;.
\ee
Even though this figure may not be accurate to within factors of ${\cal O}(10)$, it is still interesting to note that it is suppressed with respect to the tree-level coupling (\ref{topyuk}). This may have implications for example for the models presented in \cite{Dudas:2009hu} which rely on the two being equal and induce additional suppression of the non-renormalisable coupling through a small singlet vev. Of course it is plausible that for some range of geometric parameters this may hold but for this particular model, at the level of accuracy to which we can work, it seems that there is additional suppression coming from integrating out the heavy modes.

It is important to keep in mind though that there is a whole tower of massive modes which must be integrated out, whereas we have calculated just the leading contribution. Since higher mass modes also have higher powers of $z_i$ in their wavefunctions their contributions are suppressed by both the higher mass and the smaller coupling to the massless modes. We therefore expect the leading contribution to be a decent approximation. 

\subsubsection*{Hypercharge flux and $b$-$\tau$ unification}

An important piece of evidence towards grand unification is that at the GUT scale the masses of the bottom quark and tau lepton unify (see \cite{Ross:2007az} for a modern analysis). There is also approximate unification for the lighter generations. Given that hypercharge flux modifies the wavefunctions for the bottom quark and tau lepton it is important to consider whether these mass relations are maintained. Indeed it was suggested that the slight non-unification for the lighter generations may be sourced by the hypercharge flux \cite{Aparicio:2011jx}. Here we are more interested in whether the accurate unification for the heaviest generation is maintained and so calculating the appropriate coupling we find
\be
Y_{H_DQ_1D_1} = 3\times 10^{-3}\;, \;\; Y_{H_DE_1L_1} = 4\times 10^{-3}\;.
\ee
To within the accuracy we are working with this is a decent match which shows that it is possible for the two to match. However it is important to state that the two are not the same and generally in the presence of hypercharge flux the two need not match which places a constraint on the magnitude of hypercharge flux near the bottom Yukawa point.

\subsection{Exotics and Dimension five proton decay}
\label{sec:exdimfiv}

An important constraint on GUT theories comes from proton decay operators. In this section we will study dimension 5 proton decay operators that are induced by integrating out massive modes.\footnote{Note that the spectrum in table \ref{examglobspec} admits dimension 4 proton decay operators though this does not affect the use of it as a case study for dimension 5 operators. Although the model is not proposed as phenomenologically viable in itself, for this and other reasons, it is worth noting that assigning all the generations to the curve $\bf{q_4+q_5}$ will forbid these operators.} The spectrum was chosen so that it exhibits a $U(1)$ symmetry which forbids dimension 5 proton decay operators. This is a common idea in F-theory GUT model building but such a symmetry, combined with the use of hypercharge for doublet-triplet splitting, implies that there are exotics in the massless spectrum whose mass is protected by the same symmetry \cite{Marsano:2009gv,Marsano:2009wr,Dudas:2010zb,Marsano:2010sq,Dolan:2011iu,Dolan:2011aq}. In this example these exotics are marked as $E_1$ and $E_2$ in table \ref{examglobspec}. It is possible to lift these exotics by giving a vev to the singlet $X_1$ but this will then induce dimension 5 proton decay operators. There is therefore a natural tension between raising the mass of the exotics, for example to comply with gauge coupling unification, and suppressing dimension 5 proton decay \cite{Dudas:2010zb,Dolan:2011aq}. Studying this generic situation requires understanding quantitatively the coefficients of the exotics-singlet coupling and of the dimension 5 proton decay operator. An initial study in understanding the generation dependence of one of the couplings that features in the calculation, the Yukawa-like coupling of the massive Higgs triplet, was performed in \cite{Camara:2011nj}. In this section we will extended this to a full calculation of the dimension 5 operator. Although of course the results are not general, and can be modified, for example by the presence of monodromies around the up-type Yukawa, this will serve as an interesting case study of such a process. Further if we are only interested in the ratio of the coefficient of the dimension 5 operator to the coefficient of the cubic coupling used to lift the exotics the normalisation of the singlet matter curve cancels thereby increasing the accuracy of the calculation.

The dimension-five proton decay operator arises after integrating out massive states in the cubic interactions
\be
W \supset \hat{Y}_u T^{KK}_u Q_1 Q_2 + \hat{Y}_d T^{KK}_d Q_1 L_1 + \hat{Y}_X X_1 \bar{T}_u^{KK} \bar{T}_d^{KK} \;,
\ee
which, after integrating out the first massive modes, lead to the operator
\be
W \supset \frac{\hat{Y}_u\hat{Y}_d\hat{Y}_X}{M_u M_d} X_1 Q_1 Q_2 Q_1 L_1 \;.
\ee
Here $T^{KK}$ denote the massive KK-type modes, following the spectrum presented in table \ref{examlocspec}, of the Higgs triplets. Note that when integrating out the modes there are 4 possibilities according to the pairings (\ref{kkpairings}), corresponding to choosing the KK mode to be in the first or second tower for the Higgs up and Higgs down fields. All of the possibilities contribute to the higher dimensional operator but for brevity we will initially consider only the contribution coming from the choice where both $T^{KK}_u$ and $T^{KK}_d$ are in the 2nd tower (and therefore $\bar{T}^{KK}_u$ and $\bar{T}^{KK}_u$ are in the first tower. We also have the coupling of the singlet to the exotics
\be
W \supset Y_{X_1E_1E_2} X_1 E_1 E_2 \;.
\ee
We can calculate all these couplings within the local theory using the framework presented in section \ref{sec:waveover}. It is simple to check that for the exotic $E_1$ the local chirality is opposite to the global chirality which means that for consistency we have to turn on substantial local Wilson lines in its wavefunction which we do by taking $a=1$ for that mode. Taking the KK momentum in (\ref{akkk}) as $k_K=1$ gives the couplings
\bea
\hat{Y}_u &\simeq& 5 \times 10^{-4}\;, \;\; \hat{Y}_d \simeq 2 \times 10^{-4}\;, \nn \\
\hat{Y}_X &\simeq& 3 \times 10^{-4}\;, \;\;Y_{X_1E_1E_2} \simeq 5 \times 10^{-3} \;,
\eea
which implies
\be
\frac{\hat{Y}_u\hat{Y}_d\hat{Y}_X}{Y_{X_1E_1E_2}} \simeq 10^{-8} \;, \label{protexo}
\ee 
while we also have 
\be
Y_{H_DQ_1D_1}Y_{H_UQ_1U_2} \simeq 10^{-5} \;. \label{prodyuk}
\ee
An interesting thing about (\ref{protexo}) is that the proton decay operator is substantially suppressed compared to the operator that lifts the exotics, even further, by a factor of $10^{-3}$, than would be expected given the suppressed Yukawa coupling values (\ref{prodyuk}). Although this is a toy model it is encouraging to note that substantial further suppression factors are possible and could possibly alleviate the tension between proton decay and gauge coupling unification discussed above by allowing for large exotics masses.

\section{Summary}
\label{sec:summary}

In this paper we studied operator coefficients in local theories of F-theory GUTs by calculating wavefunctions and their overlaps. We solved for wavefunction profiles and overlaps in general linear Higgs and flux backgrounds for general gauge groups. We then applied these results to a model based around a point of $E_8$ enhancement and calculated coefficients of some of the operators in the theory. 

The model for which we calculated the coefficients is just a toy model used to illustrate the type of calculations that can be performed. The aim of this paper is to provide some of the tools and framework that allow for studying more generally the type of theories that arise near a point of $E_8$. The interplay between the local values of the metric, flux and Higgs field, and the local spectrum of operator coefficients is far too rich to have been systematically studied. For example, although we developed the formalism, we did not use local Wilson lines in the example model to suppress operators. Even the type of operators we studied are only a small set of those based at a point of $E_8$, and other operators can be studied using the methods and results of this paper. It would be very interesting to study more methodically the possible phenomenology that can arise from these models in future work. 

However, even within the current study, there are some general phenomenological aspects worth mentioning. It seems that it may be a challenge to recreate an ${\cal O}(1)$ top Yukawa coupling within a valid local effective theory. Indeed in the example model we found that it is too small. Although this does not have to be so always, generically the reason is that the coupling is suppressed by the size of the local patch, effectively by its modular weight under the $S_{GUT}$ modulus. We suggested that this may point towards point-like, rather than curve-like, localisation of the fields involved which can be achieved by strong flux amongst other things.\footnote{In \cite{Cecotti:2010bp} a proposal for localising point-like matter was made using local monodromies, it would be interesting to see if such localisation could be used to enhance the absolute value of the top quark coupling.} A more general phenomenological conclusion is that quite generally factors that naively may be of ${\cal O}(1)$ can easily be a few orders of magnitude smaller. A particular examples where this turned out to be useful is in calculating dimension-five proton decay in the presence of a (broken) $U(1)_{PQ}$ where we found that the non-renormalisable operator was more suppressed than expected.

A primary theme of the work presented is a study of the constraints on the local theory which turn out to be non-trivial. For example we argued that generically the local approach of using an 8-dimensional Higgsed theory to describe an intersecting brane setup, is not valid over the full $S_{GUT}$ if the normal directions to it in the CY are large, the so called decoupling limit. Indeed this was taken as a practical motivation for working in a patch within $S_{GUT}$. We also studied in detail the relation between the local values of the flux and the local massless spectrum which imposes restrictions on both.

There are a number of phenomenologically motivated improvements that can be made to the simplest models studied in this paper. An important one is the incorporation of local monodromies, perhaps using the tools presented in \cite{Cecotti:2010bp}. This is likely to require numerical methods in solving for the wavefunctions since complete analytic solutions to the type of differential equations that arise in this case are not known.

Perhaps more radically it may be that the size of operator coefficients, or natural global embeddings, require that we go to regions of parameter space that are beyond the validity of the local theory itself. In particular this applies to the case where the component of the metric of the normal direction to the brane is larger than the parallel ones. In that case a more appropriate theory would seem to be a 10-dimensional one with a product gauge group, ie. separating the bulk branes out rather than treating them as small deformations of the GUT brane. It is not clear how much it would be possible to calculate in such a theory for a general bulk geometry and whether the local approach can still be applied. On a similar theme it would be interesting to see if there is some sense of a local theory that can be applied to Heterotic constructions where the localisation of the wavefunctions would be done purely by flux effects. In particular if it is possible to localise using fluxes onto the analogue of a point of $E_8$ perhaps using the line-bundle constructions of \cite{Anderson:2011ns,Anderson:2012yf}. 

As discussed in the introduction the local approach to model building has a number of advantages, in particular to do with calculability and generality within the landscape. In this work we have taken some steps towards this aim for a simple class of models. There is much work to be done if we are able to reach the full potential of a phenomenologically viable local model which recreates, at least approximately, the operator coefficients that we observe and constrain in experiments. And of course it may be that the real world does not correspond to a local model at all. But it seems to be a direction worth pursuing in that it has the potential of forming a relatively direct route to quantitative comparison of a large and general class of models with a large experimental dataset.

\begin{center}\subsection*{Acknowledgments}\end{center}
I would like to thank Pablo Camara, Emilian Dudas and Fernando Marchesano for
very useful comments and discussions. The research of EP is supported by a Marie Curie  Intra European
Fellowship within the 7th European Community Framework Programme.
\vspace*{1.0cm}

\appendix

\section{Semi-local constraints on matter spectrum}

In this appendix we calculate some constraints on the restriction of hypercharge flux coming from the embedding of the matter curves into the full $S_{GUT}$. Given a spectrum of fields within a local patch on $S_{GUT}$ as discussed in section \ref{sec:locspe}, the constraints on the full $S_{GUT}$ derived in this section can be used to understand what the spectrum away from the local patch must be like in order to be compatible.

The techniques used to derive the homology classes of, and therefore the possible flux restrictions to, the matter curves are described in detail in \cite{Dudas:2010zb}, to which we refer for further details and here we will simply state the results. Following precisely the methodology and notation of \cite{Dudas:2010zb} we start from a fully split spectral cover
\be
C_{10} = \left(a_1 V + a_6 U \right) \left(a_2 V + a_7 U \right) \left(a_3 V + a_8 U\right) \left(a_4 V + a_9 U\right) \left(a_5 V + a_{10} U\right) = 0 \;. \label{spect10z2}
\ee
Here the $a_I$ are some as yet undetermined coefficients that are functions on $S$. The spectral cover can also be written as
\be
C_{10} = b_0 U^5 + b_2 V^2 U^3 + b_3 V^3 U^2 + b_4 V^4 U + b_5 V^5 = 0 \;, \label{spect10}
\ee
where we know that the $b_i$ are zero sections of the bundle $\eta - i c_1$ \cite{Donagi:2009ra,Marsano:2009gv}. Here $c_1$ is the first Chern class of the tangent bundle of $S_{GUT}$ and $\eta = 6 c_1 - t$ with $-t$ being the first Chern class of the normal bundle to $S_{GUT}$. Comparing (\ref{spect10}) and (\ref{spect10z2}) then implies that the $a_I$ are sections of bundles as shown in table \ref{tab:aIsect2111}.
\begin{table}
\centering
\begin{tabular}{|c|c|}
\hline
Section & $c_1$(Bundle)\\
\hline
$a_1$ & $\eta - c_1 - \tilde{\chi}$ \\
\hline
$a_2$ & $- c_1 + \chi_7$ \\
\hline
$a_3$ & $-c_1 + \chi_8$\\
\hline
$a_4$ & $-c_1 + \chi_9$ \\
\hline
$a_5$ & $-c_1 + \chi_{10}$\\
\hline
$a_6$ & $\eta - \tilde{\chi}$\\
\hline
$a_7$ & $\chi_7$ \\
\hline
$a_8$ & $\chi_8$\\
\hline
$a_9$ & $\chi_9$\\
\hline
$a_{10}$ & $\chi_{10}$\\
\hline
\end{tabular}
\caption{Table showing the first Chern classes of the line bundles that the $a_I$ are sections of. The forms $\chi_{\{7,8,9,10\}}$ are unspecified and we define $\tilde{\chi}=\chi_7 + \chi_8 + \chi_9 + \chi_{10}$.}
\label{tab:aIsect2111}
\end{table}
Here $\chi_{\{7,8,9,10\}}$ are unspecified and we define $\tilde{\chi}=\chi_7 + \chi_8 + \chi_9+\chi_{10}$.

To obtain the homology classes of the matter curves we need to solve the constraint $b_1=0$ which we do by setting
\bea
a_1 &=& - \left( a_2 a_8 a_9 a_{10} + a_3 a_7 a_9 a_{10} + a_4 a_7 a_8 a_{10} + a_5 a_7 a_8 a_{9} \right) \;, \nn \\
a_6 &=& a_7 a_8 a_9 a_{10}\;, \label{b1sol2111}
\eea
which also imposes the homology constraint
\be
\eta = 2 \tilde{\chi}\;. \label{homconst}
\ee
With this choice it can be checked that there are no exotic non-Kodaira singularities on $S_{GUT}$ over any of the curves $a_i$.\footnote{Note that the relations (\ref{b1sol2111}) solve the constraint $b_1=0$ also with an additional arbitrary section $c$ multiplying each of the right-hand sides. However if $c$ has a zero section then it induces a non-Kodaira type singularity on $S_{GUT}$ and so we should set $c=1$ which implies the form (\ref{homconst}).} 

The matter curves are given by the equations
\bea
P_{10} &=& b_5 =0\;, \\
P_5 &=& b_3^2 b_4 - b_2 b_3 b_5 + b_0 b_5^2 =0\;. \label{p5p10}
\eea
Which, after imposing (\ref{b1sol2111}), give the curve homology classes as specified in table \ref{tab:curves2111}.
\begin{table}[ht]
\center
{\small
\begin{tabular}{|cccccc|}
\hline
Matter & Charge & Equation & Homology & $N$ & $M_{U(1)}$ \\
\hline
$\f_{1}$    & $-{\bf q}_1-{\bf q}_2$ & $a_3 a_9 a_{10} + a_4 a_8 a_{10} + a_5 a_8 a_{9}$     & $-c_1 + \chi_8 + \chi_9 + \chi_{10}$  & $-N_7$  & $M_{\f_1}$ \\
$\f_{2}$    & $-{\bf q}_1-{\bf q}_3$ & $a_2 a_9 a_{10} + a_4 a_7 a_{10} + a_5 a_7 a_{9}$     & $-c_1 + \chi_7 + \chi_9 + \chi_{10}$  & $-N_8$  & $M_{\f_2}$ \\
$\f_{3}$    & $-{\bf q}_1-{\bf q}_4$ & $a_2 a_8 a_{10} + a_3 a_7 a_{10} + a_5 a_7 a_{8}$     & $-c_1 + \chi_7 + \chi_8 + \chi_{10}$  & $-N_9$  & $M_{\f_3}$ \\
$\f_{4}$    & $-{\bf q}_1-{\bf q}_5$ & $a_2 a_8 a_{9}  + a_3 a_7 a_{9}  + a_4 a_7 a_{8}$     & $-c_1 + \chi_7 + \chi_8 + \chi_9$     & $-N_{10}$     & $M_{\f_4}$ \\
$\f_{5}$    & $-{\bf q}_2-{\bf q}_3$ & $a_2 a_8 + a_3 a_7$                       						 & $-c_1 + \chi_7 + \chi_8$      & $N_7+N_8$                        & $M_{\f_5}$ \\
$\f_{6}$    & $-{\bf q}_2-{\bf q}_4$ & $a_2 a_9 + a_4 a_7$                       						 & $-c_1 + \chi_7 + \chi_9$      & $N_7+N_9$                        & $M_{\f_6}$ \\
$\f_{7}$    & $-{\bf q}_2-{\bf q}_5$ & $a_2 a_{10} + a_5 a_7$								     & $-c_1 + \chi_7 + \chi_{10}$  & $N_7+N_{10}$  & $M_{\f_7}$ \\
$\f_{8}$    & $-{\bf q}_3-{\bf q}_4$ & $a_3 a_9 + a_4 a_8$									     & $-c_1 + \chi_8 + \chi_9$  	  & $N_8+N_9$  		& $M_{\f_8}$ \\
$\f_{9}$    & $-{\bf q}_3-{\bf q}_5$ & $a_3 a_{10} + a_5 a_8$								     & $-c_1 + \chi_8 + \chi_{10}$  & $N_8+N_{10}$  & $M_{\f_9}$ \\
$\f_{10}$   & $-{\bf q}_4-{\bf q}_5$ & $a_4 a_{10} + a_5 a_9$								     & $-c_1 + \chi_9 + \chi_{10}$  & $N_9+N_{10}$  & $M_{\f_{10}}$ \\
$\te_{1}$   & ${\bf q}_1$      & $a_1$                                     & $- c_1+\tilde{\chi}$   & $0$ & $M_{\te_1}$ \\
$\te_{2}$   & ${\bf q}_2$      & $a_2$                                     & $-c_1 + \chi_7$               & $N_7$                            & $M_{\te_2}$ \\
$\te_{3}$   & ${\bf q}_3$      & $a_3$                                     & $-c_1 + \chi_8$               & $N_8$                            & $M_{\te_3}$ \\
$\te_{4}$   & ${\bf q}_4$      & $a_4$                                     & $-c_1 + \chi_9$               & $N_9$                            & $M_{\te_4}$ \\
$\te_{5}$   & ${\bf q}_5$      & $a_5$                                     & $-c_1 + \chi_{10}$            & $N_{10}$                         & $M_{\te_5}$ \\
\hline
\end{tabular}
}
\caption{Table showing curves and flux restrictions on $S_{GUT}$ in the absence of monodromies. The fluxes are constrained so that $N_7+N_8+N_9+N_{10}=0$.}
\label{tab:curves2111}
\end{table}
The table also shows the possible flux restrictions over $S_{GUT}$ for the matter curves which in turn constrain the possible chiral spectrum. Given the flux values $N_i$ and $M_i$ the resulting spectrum is determined as
\bea
n_{(3,1)_{-1/3}} - n_{(\bar{3},1)_{+1/3} } &=& M_{5} \;, \nn \\
n_{(1,2)_{+1/2}} - n_{(1,2)_{-1/2} } &=& M_{5} + N \;,
\eea
for the 5-matter curves and
\bea
n_{(3,2)_{+1/6}} - n_{(\bar{3},2)_{-1/6} } &=& M_{10} \;, \nn \\
n_{(\bar{3},1)_{-2/3}} - n_{(3,1)_{+2/3} } &=& M_{10} - N \;, \nn \\
n_{(1,1)_{+1}} - n_{(1,1)_{-1} } &=& M_{10} + N\;,
\eea
for the 10-matter curves.\footnote{It is worth noting when comparing the local and global values of the fluxes that the integrated flux values here $M$ and $N$ are combinations of the hypercharge and $U(1)$ fluxes such that $M_{\bf 5} = M^{U(1)} - \frac13 N^Y$, $M_{\bf 10} = M^{U(1)} + \frac16 N^Y$, $N = \frac56 N^Y$.}

The example model of section \ref{sec:modbui} is realised explicitly by taking the non-vanishing flux values presented in table \ref{examflux}.
\begin{table}
\centering
\begin{tabular}{|c|c|}
\hline
Flux & Value  \\
\hline
$M_{\f_6}$ & +1\\
\hline
$M_{\f_{2}}$ & -2\\
\hline
$M_{\f_{10}}$ & -2\\
\hline
$M_{\te_1}$ & +1\\
\hline
$M_{\te_2}$ & +3\\
\hline
$M_{\te_5}$ & -1\\
\hline
$N_7$ & -1\\
\hline
$N_{10}$ & +1\\
\hline
\end{tabular}
\caption{Table showing the flux values used to obtain the spectrum of the example model in section \ref{sec:modbui}.}
\label{examflux}
\end{table}

\end{document}